\documentstyle[10pt,epsf,epsfig,hangcaption,xspace,amssymb,amsfonts,amsmath,amsthm,cite,
dp_delphititle,subfigure]{dp_delphi}

\makeindex
\def\DpPaperGroup{EP-PH}
\def\DpPaperRef{2004-066}
\def\DpDate{21 September 2004}
\def\DpAuthors{DELPHI Collaboration}
\def\DpSubmit{(Euro. Phys. J.C44 (2005) 147-159)}
\def\DpTitle{
Flavour Independent Searches \\ for Hadronically Decaying \\ Neutral Higgs Bosons}
\def\DpComment{ }
\def\DpEMail{ }

\newcommand{\epem}{\mbox{${\mathrm{e^+e^-}}$}}

\newcommand{\hZ}{\mbox{${\mathrm{hZ}}$}}
\newcommand{\hA}{\mbox{${\mathrm{hA}}$}}
\newcommand{\ZZ}{\mbox{${\mathrm{ZZ}}$}}

\newcommand{\gammagamma}{\mbox{${\mathrm{\gamma \gamma}}$}}
\newcommand{\Zg}{\mbox{${\mathrm{Z\gamma}}$}}
\newcommand{\Zgstar}{\mbox{${\mathrm{Z\gamma^*}}$}}

\newcommand{\qcd}{\mbox{${\mathrm{q\bar{q}(\gamma)}}$}}

\newcommand{\bdqq}{\mbox{${\mathrm{q\bar{q}}}$}}
\newcommand{\bdss}{\mbox{${\mathrm{s\bar{s}}}$}}
\newcommand{\bdcc}{\mbox{${\mathrm{c\bar{c}}}$}}
\newcommand{\bdbb}{\mbox{${\mathrm{b\bar{b}}}$}}
\newcommand{\bdgg}{\mbox{${\mathrm{gg}}$}}

\newcommand{\hnunu}{\mbox{${\mathrm{h\nu\bar{\nu}}}$}}

\newcommand{\Wenu}{\mbox{${\mathrm{We\nu}}$}}

\newcommand{\qqnunu}{\mbox{${\mathrm{\nu\bar{\nu}q\bar{q}}}$}}
\newcommand{\qqmumu}{\mbox{${\mathrm{q\bar{q}\mu^+\mu^-}}$}}
\newcommand{\qqee}{\mbox{${\mathrm{q\bar{q}e^+e^-}}$}}

\newcommand{\Z}{\mbox{$\mathrm{Z}$}}
\newcommand{\h}{\mbox{$\mathrm{h}$}}

\newcommand{\mh}{\mbox{$m_{\mathrm{h}}$}}
\newcommand{\mA}{\mbox{$m_{\mathrm{A}}$}}
\newcommand{\mZ}{\mbox{$m_{\mathrm{Z}}$}}

\newcommand{\chz}{\mbox{$C^{2}_{\mathrm{hZ}}$}}
\newcommand{\rhz}{\mbox{$R_{\mathrm{hZ}}$}}
\newcommand{\cha}{\mbox{$C^{2}_{\mathrm{hA}}$}}
\newcommand{\rha}{\mbox{$R_{\mathrm{hA}}$}}

\newcommand{\ra}{$\rightarrow$}
\newcommand{\massunit}{$\mathrm{GeV/}c^2$}

\def\NPB#1#2#3{Nucl.~Phys. {\bf B#1} (#2) #3}
\def\PLB#1#2#3{Phys.~Lett. {\bf B#1} (#2) #3}
\def\PRD#1#2#3{Phys.~Rev. {\bf D#1} (#2) #3}

\def\ZPC#1#2#3{Z.~Phys. {\bf C#1} (#2) #3}

\def\CPC#1#2#3{Comp.~Phys.~Comm. {\bf #1} (#2) #3}
\def\NIMA#1#2#3{Nucl.~Instr.~and~Meth. {\bf A#1} (#2) #3} 
\def\EPJ#1#2#3{Eur.~Phys.~J. {\bf C#1} (#2) #3} 

\begin{document}
\makeatletter
\input{coll.sty}
\makeatother

\begin{titlepage}
\pagenumbering{roman}

\CERNpreprint{\DpPaperGroup}{\DpPaperRef}   
\date{{\small\DpDate}}                      
\title{\DpTitle}                            
\address{\DpAuthors}                        

\begin{shortabs}                            
\noindent
This paper describes flavour independent searches for hadronically decaying neutral Higgs bosons
in the data collected by the DELPHI experiment at LEP, at centre-of-mass energies between 189 and 209 GeV.
The collected data-set corresponds to an integrated luminosity of around 
610 pb$^{-1}$. The \epem\ra~\hA\ and \epem\ra~\hZ\ processes are considered, with direct Higgs boson decays into hadrons. 
No evidence for Higgs boson production is found, and cross-section limits
are set as a function of the Higgs boson masses. No explicit assumptions are made on the underlying 
physics beyond the Standard Model, allowing interpretation of the data in a large class of models.
\end{shortabs}

\vfill

\begin{center}
\DpSubmit \ \\          
\DpComment \ \\
\DpEMail \ \\
\end{center}

\vfill
\clearpage

\headsep 10.0pt

\addtolength{\textheight}{10mm}
\addtolength{\footskip}{-5mm}
\begingroup
%
\newcommand{\DpName}[2]{\hbox{#1$^{\ref{#2}}$},\hfill}
\newcommand{\DpNameTwo}[3]{\hbox{#1$^{\ref{#2},\ref{#3}}$},\hfill}
\newcommand{\DpNameThree}[4]{\hbox{#1$^{\ref{#2},\ref{#3},\ref{#4}}$},\hfill}
\newskip\Bigfill \Bigfill = 0pt plus 1000fill
\newcommand{\DpNameLast}[2]{\hbox{#1$^{\ref{#2}}$}\hspace{\Bigfill}}

%
\footnotesize
\noindent
\DpName{J.Abdallah}{LPNHE}
\DpName{P.Abreu}{LIP}
\DpName{W.Adam}{VIENNA}
\DpName{P.Adzic}{DEMOKRITOS}
\DpName{T.Albrecht}{KARLSRUHE}
\DpName{T.Alderweireld}{AIM}
\DpName{R.Alemany-Fernandez}{CERN}
\DpName{T.Allmendinger}{KARLSRUHE}
\DpName{P.P.Allport}{LIVERPOOL}
\DpName{U.Amaldi}{MILANO2}
% ...
\normalsize
\endgroup

\titlefoot{Department of Physics and Astronomy, Iowa State
     University, Ames IA 50011-3160, USA
    \label{AMES}}
\titlefoot{Physics Department, Universiteit Antwerpen,
     Universiteitsplein 1, B-2610 Antwerpen, Belgium \\
     \indent~~and IIHE, ULB-VUB,
     Pleinlaan 2, B-1050 Brussels, Belgium \\
     \indent~~and Facult\'e des Sciences,
     Univ. de l'Etat Mons, Av. Maistriau 19, B-7000 Mons, Belgium
    \label{AIM}}
\titlefoot{Physics Laboratory, University of Athens, Solonos Str.
     104, GR-10680 Athens, Greece
    \label{ATHENS}}
\titlefoot{Department of Physics, University of Bergen,
     All\'egaten 55, NO-5007 Bergen, Norway
    \label{BERGEN}}
\titlefoot{Dipartimento di Fisica, Universit\`a di Bologna and INFN,
     Via Irnerio 46, IT-40126 Bologna, Italy
    \label{BOLOGNA}}
\titlefoot{Centro Brasileiro de Pesquisas F\'{\i}sicas, rua Xavier Sigaud 150,
     BR-22290 Rio de Janeiro, Brazil \\
     \indent~~and Depto. de F\'{\i}sica, Pont. Univ. Cat\'olica,
     C.P. 38071 BR-22453 Rio de Janeiro, Brazil \\
     \indent~~and Inst. de F\'{\i}sica, Univ. Estadual do Rio de Janeiro,
     rua S\~{a}o Francisco Xavier 524, Rio de Janeiro, Brazil
    \label{BRASIL}}
\titlefoot{Coll\`ege de France, Lab. de Physique Corpusculaire, IN2P3-CNRS,
     FR-75231 Paris Cedex 05, France
    \label{CDF}}
\titlefoot{CERN, CH-1211 Geneva 23, Switzerland
    \label{CERN}}
\titlefoot{Institut de Recherches Subatomiques, IN2P3 - CNRS/ULP - BP20,
     FR-67037 Strasbourg Cedex, France
    \label{CRN}}
\titlefoot{Now at DESY-Zeuthen, Platanenallee 6, D-15735 Zeuthen, Germany
    \label{DESY}}
\titlefoot{Institute of Nuclear Physics, N.C.S.R. Demokritos,
     P.O. Box 60228, GR-15310 Athens, Greece
    \label{DEMOKRITOS}}
\titlefoot{FZU, Inst. of Phys. of the C.A.S. High Energy Physics Division,
     Na Slovance 2, CZ-180 40, Praha 8, Czech Republic
    \label{FZU}}
\titlefoot{Dipartimento di Fisica, Universit\`a di Genova and INFN,
     Via Dodecaneso 33, IT-16146 Genova, Italy
    \label{GENOVA}}
\titlefoot{Institut des Sciences Nucl\'eaires, IN2P3-CNRS, Universit\'e
     de Grenoble 1, FR-38026 Grenoble Cedex, France
    \label{GRENOBLE}}
\titlefoot{Helsinki Institute of Physics, P.O. Box 64,
     FIN-00014 University of Helsinki, Finland
    \label{HELSINKI}}
\titlefoot{Joint Institute for Nuclear Research, Dubna, Head Post
     Office, P.O. Box 79, RU-101 000 Moscow, Russian Federation
    \label{JINR}}
\titlefoot{Institut f\"ur Experimentelle Kernphysik,
     Universit\"at Karlsruhe, Postfach 6980, DE-76128 Karlsruhe,
     Germany
    \label{KARLSRUHE}}
\titlefoot{Institute of Nuclear Physics PAN,Ul. Radzikowskiego 152,
     PL-31142 Krakow, Poland
    \label{KRAKOW1}}
\titlefoot{Faculty of Physics and Nuclear Techniques, University of Mining
     and Metallurgy, PL-30055 Krakow, Poland
    \label{KRAKOW2}}
\titlefoot{Universit\'e de Paris-Sud, Lab. de l'Acc\'el\'erateur
     Lin\'eaire, IN2P3-CNRS, B\^{a}t. 200, FR-91405 Orsay Cedex, France
    \label{LAL}}
\titlefoot{School of Physics and Chemistry, University of Lancaster,
     Lancaster LA1 4YB, UK
    \label{LANCASTER}}
\titlefoot{LIP, IST, FCUL - Av. Elias Garcia, 14-$1^{o}$,
     PT-1000 Lisboa Codex, Portugal
    \label{LIP}}
\titlefoot{Department of Physics, University of Liverpool, P.O.
     Box 147, Liverpool L69 3BX, UK
    \label{LIVERPOOL}}
\titlefoot{Dept. of Physics and Astronomy, Kelvin Building,
     University of Glasgow, Glasgow G12 8QQ
    \label{GLASGOW}}
\titlefoot{LPNHE, IN2P3-CNRS, Univ.~Paris VI et VII, Tour 33 (RdC),
     4 place Jussieu, FR-75252 Paris Cedex 05, France
    \label{LPNHE}}
\titlefoot{Department of Physics, University of Lund,
     S\"olvegatan 14, SE-223 63 Lund, Sweden
    \label{LUND}}
\titlefoot{Universit\'e Claude Bernard de Lyon, IPNL, IN2P3-CNRS,
     FR-69622 Villeurbanne Cedex, France
    \label{LYON}}
\titlefoot{Dipartimento di Fisica, Universit\`a di Milano and INFN-MILANO,
     Via Celoria 16, IT-20133 Milan, Italy
    \label{MILANO}}
\titlefoot{Dipartimento di Fisica, Univ. di Milano-Bicocca and
     INFN-MILANO, Piazza della Scienza 2, IT-20126 Milan, Italy
    \label{MILANO2}}
\titlefoot{IPNP of MFF, Charles Univ., Areal MFF,
     V Holesovickach 2, CZ-180 00, Praha 8, Czech Republic
    \label{NC}}
\titlefoot{NIKHEF, Postbus 41882, NL-1009 DB
     Amsterdam, The Netherlands
    \label{NIKHEF}}
\titlefoot{National Technical University, Physics Department,
     Zografou Campus, GR-15773 Athens, Greece
    \label{NTU-ATHENS}}
\titlefoot{Physics Department, University of Oslo, Blindern,
     NO-0316 Oslo, Norway
    \label{OSLO}}
\titlefoot{Dpto. Fisica, Univ. Oviedo, Avda. Calvo Sotelo
     s/n, ES-33007 Oviedo, Spain
    \label{OVIEDO}}
\titlefoot{Department of Physics, University of Oxford,
     Keble Road, Oxford OX1 3RH, UK
    \label{OXFORD}}
\titlefoot{Dipartimento di Fisica, Universit\`a di Padova and
     INFN, Via Marzolo 8, IT-35131 Padua, Italy
    \label{PADOVA}}
\titlefoot{Rutherford Appleton Laboratory, Chilton, Didcot
     OX11 OQX, UK
    \label{RAL}}
\titlefoot{Dipartimento di Fisica, Universit\`a di Roma II and
     INFN, Tor Vergata, IT-00173 Rome, Italy
    \label{ROMA2}}
\titlefoot{Dipartimento di Fisica, Universit\`a di Roma III and
     INFN, Via della Vasca Navale 84, IT-00146 Rome, Italy
    \label{ROMA3}}
\titlefoot{DAPNIA/Service de Physique des Particules,
     CEA-Saclay, FR-91191 Gif-sur-Yvette Cedex, France
    \label{SACLAY}}
\titlefoot{Instituto de Fisica de Cantabria (CSIC-UC), Avda.
     los Castros s/n, ES-39006 Santander, Spain
    \label{SANTANDER}}
\titlefoot{Inst. for High Energy Physics, Serpukov
     P.O. Box 35, Protvino, (Moscow Region), Russian Federation
    \label{SERPUKHOV}}
\titlefoot{J. Stefan Institute, Jamova 39, SI-1000 Ljubljana, Slovenia
     and Laboratory for Astroparticle Physics,\\
     \indent~~Nova Gorica Polytechnic, Kostanjeviska 16a, SI-5000 Nova Gorica, Slovenia, \\
     \indent~~and Department of Physics, University of Ljubljana,
     SI-1000 Ljubljana, Slovenia
    \label{SLOVENIJA}}
\titlefoot{Fysikum, Stockholm University,
     Box 6730, SE-113 85 Stockholm, Sweden
    \label{STOCKHOLM}}
\titlefoot{Dipartimento di Fisica Sperimentale, Universit\`a di
     Torino and INFN, Via P. Giuria 1, IT-10125 Turin, Italy
    \label{TORINO}}
\titlefoot{INFN,Sezione di Torino, and Dipartimento di Fisica Teorica,
     Universit\`a di Torino, Via P. Giuria 1,\\
     \indent~~IT-10125 Turin, Italy
    \label{TORINOTH}}
\titlefoot{Dipartimento di Fisica, Universit\`a di Trieste and
     INFN, Via A. Valerio 2, IT-34127 Trieste, Italy \\
     \indent~~and Istituto di Fisica, Universit\`a di Udine,
     IT-33100 Udine, Italy
    \label{TU}}
\titlefoot{Univ. Federal do Rio de Janeiro, C.P. 68528
     Cidade Univ., Ilha do Fund\~ao
     BR-21945-970 Rio de Janeiro, Brazil
    \label{UFRJ}}
\titlefoot{Department of Radiation Sciences, University of
     Uppsala, P.O. Box 535, SE-751 21 Uppsala, Sweden
    \label{UPPSALA}}
\titlefoot{IFIC, Valencia-CSIC, and D.F.A.M.N., U. de Valencia,
     Avda. Dr. Moliner 50, ES-46100 Burjassot (Valencia), Spain
    \label{VALENCIA}}
\titlefoot{Institut f\"ur Hochenergiephysik, \"Osterr. Akad.
     d. Wissensch., Nikolsdorfergasse 18, AT-1050 Vienna, Austria
    \label{VIENNA}}
\titlefoot{Inst. Nuclear Studies and University of Warsaw, Ul.
     Hoza 69, PL-00681 Warsaw, Poland
    \label{WARSZAWA}}
\titlefoot{Fachbereich Physik, University of Wuppertal, Postfach
     100 127, DE-42097 Wuppertal, Germany
    \label{WUPPERTAL}}
\addtolength{\textheight}{-10mm}
\addtolength{\footskip}{5mm}
\clearpage

\headsep 30.0pt
\end{titlepage}

%
\pagenumbering{arabic}                              
\setcounter{footnote}{0}                            %
\large
\section{Introduction}
\label{sec:introduction}

As long as bosonic decay channels are kinematically closed, the Standard Model (SM) Higgs boson 
decays preferentially into the heaviest accessible fermion pair. This leads to the dominance of 
Higgs boson decays to b quarks in most of the mass range accessible at LEP. In extensions of the 
SM, however, the Higgs boson couplings to b quarks might well be suppressed. This can for example occur 
in the Minimal Supersymmetric Standard Model (MSSM)~\cite{MSSM-flavind},
and in general Two Higgs Doublet Models (2HDM)~\cite{2HDM-flavind}. Suppressed couplings to b-quarks are also possible 
in models with composite Higgs and gauge bosons, where the dominant Higgs boson decay channel 
is gluonic~\cite{calmet}.  The SM Higgs boson search at LEP strongly relies on the identification of 
b~quarks or $\tau$~leptons to separate a possible signal from background processes, and has reduced sensitivity 
to the final states predicted by such alternative models. 

This paper presents searches that only assume direct hadronic decays of the Higgs bosons. The scope of these 
analyses is thus more general, and encompasses the models referred to above.
However, rather than stating the results in explicit theoretical frameworks, model independent bounds on production 
cross-sections are derived, allowing to confront a wide class of models with the data.

Two Higgs boson production processes are studied in this paper: the Higgsstrahlung 
process \epem\ra~\hZ, and Higgs boson pair production \epem\ra~\hA, which is present in all extensions 
of the Standard Model Higgs sector beyond one doublet. Although the notation is familiar from
CP conserving models, the CP properties of h and A are not necessarily specified here.

Flavour independent searches for hadronic Higgs boson decays have been published by the other LEP 
collaborations~\cite{alephflb,l3flb,opalflb}.

The paper is organized as follows. The common framework and search strategy is described in 
Section~\ref{sec:generalstrategy}, and is followed by an overview of the experimental setup and the 
data sets in Section~\ref{sec:dataandsimulation}. 
The search for hadronic decays of hA pairs is described in Section~\ref{sec:hAproduction}. 
Section~\ref{sec:hZproduction} describes the search for hZ production, where the Higgs boson is accompanied by jets,
missing energy, electrons, or muons from the Z boson decay. Section~\ref{sec:results} combines the various analyses, and gives 
flavour independent results on the hZ and hA cross-sections, as a function of the Higgs boson 
masses. Section~\ref{sec:summary} concludes the paper.


\section{General features}
\label{sec:generalstrategy}

The analyses presented in this paper rely essentially on kinematic and event shape information. Without the 
background rejection allowed by b quark or $\tau$ lepton identification, the signal contamination 
by quark pair production and/or vector boson pair production can be important, 
and dijet mass information must be used to discriminate signal from background. The way in which this information is 
exploited depends on the search channel, and is detailed in the corresponding sections.


The search results independence from the hadronic flavour of the Higgs bosons decay products is obtained by considering 
only the decay mode with lowest sensitivity in each search channel. The precise procedure varies again from one channel to
the other, and details are given for each analysis in the corresponding description.

All search results presented in this work are interpreted using a
modified frequentist technique based on the extended
likelihood ratio \cite{alrmc}. For a given experiment, the test
statistic $Q$ is defined as the likelihood ratio of the signal+background hypothesis (s+b) to
the background hypothesis (b), computed from the number of observed and expected events
in both hypotheses. 
Measured discriminating variables can be used to assign a signal to background ratio to individual events.
Probability density functions (PDFs) for $Q$ in the b and s+b hypotheses are
built using Monte Carlo sampling of the expected background and signal rates, and of the optional discriminating
variable distributions. The confidence levels $\mathrm{CL_b}$ and
$\mathrm{CL_{s+b}}$ are defined as the integrals of the b and s+b
PDFs for $Q$ between $-\infty$ and the observed value
$Q_{obs}$. 

A signal hypothesis to which there is no sensitivity might be uncorrectly excluded by $\mathrm{CL_{s+b}}$ in case a 
downward fluctuation of the data occurs. To prevent this, we adopt the conservative approach of defining
the confidence level in the signal hypothesis, $\mathrm{CL_s}$, as the ratio $\mathrm{CL_{s+b}/CL_b}$.
1-$\mathrm{CL_s}$ measures the confidence
with which the signal hypothesis can be rejected, and will be larger
than 0.95 for an exclusion confidence of 95\%.

When no signal is found, upper bounds on the production cross-section times the branching 
fraction into hadrons are extracted for both production processes as a function of the Higgs 
boson masses. These bounds are expressed in terms of reference cross-sections, defined 
as follows. The production rate of any final state resulting from \epem\ra~\hZ\ can be expressed 
in terms of the SM hZ cross-section, $\sigma_{{\mathrm hZ}}^{{\mathrm SM}}$ \cite{gkwsig}. 
The reference cross-section for \epem\ra~\hA, $\sigma_{{\mathrm hA}}^{{\mathrm ref}}$, is 
calculated with {\tt HZHA} \cite{hzha}, assuming the 2HDM and the absence of any mixing in the Higgs sector.
The result depends only on electroweak parameters and the h and A Higgs boson masses, and remains valid for models with
any number of Higgs doublets or singlets.
Suppression terms coming from branching fractions and possible 
suppression of the Higgs boson couplings to the Z (such factors are hereafter denoted by $BR$ 
and $R$, respectively) all factorize, and are hidden in a generic (model independent) suppression 
factor :

\begin{eqnarray*}
\sigma_{{\mathrm hZ \rightarrow Z+hadrons}}   &=&    \sigma_{{\mathrm hZ}}^{{\mathrm SM}}  \times \rhz
                                                         \times BR({\mathrm h\rightarrow hadrons}) \\
                              &\equiv& \sigma_{{\mathrm hZ}}^{{\mathrm SM}}  \times \chz;\\
\sigma_{{\mathrm hA \rightarrow hadrons}}     &=&    \sigma_{{\mathrm hA}}^{{\mathrm ref}} \times \rha
                                                         \times BR({\mathrm h\rightarrow hadrons})
                                                         \times BR({\mathrm A\rightarrow hadrons}) \\
                              &\equiv& \sigma_{{\mathrm hA}}^{{\mathrm ref}} \times \cha.\\
\end{eqnarray*}

\noindent Our results will be expressed in terms of \chz\ and \cha.

Although the traditional CP-conserving Higgs boson nomenclature is used throughout the paper, part of the results
can also be extended to the context of CP-violating Higgs sectors. For \hA~production, the parity and 
charge conjugation properties of the Higgs bosons do not play any role in the decay kinematics. The results 
on \cha\ are therefore universally valid for the pair production of any two 
Higgs bosons (i.e. $\mathrm{h_{i}h_{j}\rightarrow hadrons}$), regardless of their CP properties. On the contrary, 
our results on \epem\ra~\hZ\ assume standard quantum numbers for the 
Higgs boson. Non-standard Higgs boson parity properties would imply different polarization of the associated 
\Z~boson, thereby affecting the angular distributions of the bosons and the selection efficiencies. Our results 
for this channel should therefore be used with care.

If kinematically open and allowed by the Higgs boson quantum numbers, cascade decays like h\ra AA may become dominant. 
The efficiency of the analyses developed below to these final states has not been evaluated.

The study also applies to non-Higgs scalar particles. The cross-sections and analyses presented here however assume 
that the produced scalars have negligible width (less than about 1~\massunit).

\section{Detector, data samples and simulation}
\label{sec:dataandsimulation}
A detailed description of the DELPHI detector and its performance during the LEP1 data taking period can be found 
in \cite{delphiperformance}. For LEP2, the vertex detector was upgraded \cite{delphivertexdetector}, 
and scintillator counters were added in regions without electromagnetic calorimetry (at 40 and 90 degrees 
from the beam axis), improving the detection of events with radiative photons in these regions.
The detector was operated in nominal conditions until September 2000, when one of the twelve sectors of
DELPHI's main tracking device underwent an irremediable failure. A special algorithm, based on
complementary subdetectors, allowed to limit the tracking efficiency loss in this region.

The data samples used in this paper were collected during the last 3 years of LEP operation (1998 to 2000), 
and are clustered around seven centre-of-mass energy values, as shown in Table~\ref{tab:luminosityLEPII}. 
The total used luminosity amounts to 610.4 pb$^{-1}$. The \hZ~analysis uses the data from all listed sets, 
while the hA search uses all but the two sets with lowest luminosity.

\begin{table}[htbp]
  \begin{center}
    \begin{tabular}{ c | c | c c c c | c c }
      \hline
      \hline
      year    & 1998 &  \multicolumn{4}{c|}{1999}  & \multicolumn{2}{c}{2000}   \\
      \hline
      $\sqrt{s}$ (GeV)       & 188.6  & 191.6  & 195.5  & 199.5  &  201.6  & 205.0  &  206.5\\
      ${\cal L}$ (pb$^{-1}$) & 158.0  & 25.9   & 76.9   & 84.3   &  41.1   &  82.0  &  142.2  \\
      \hline
    \end{tabular}
  \end{center}
  \vspace{-1em}
  \caption{Centre-of-mass energy and integrated luminosity of the data collected by the DELPHI detector in the years 1998-2000.}
  \label{tab:luminosityLEPII}
\end{table}

The simulation samples used for this study have been produced at the centre-of-mass energies given in Table~\ref{tab:luminosityLEPII};
details are given below. In all cases, samples have been processed that comply with the 
experimental situation before and after the detector failure mentioned above.

For the hZ channel, two-fermion background events are simulated using the 
{\tt KK2F} \cite{KK2F} generator, while {\tt WPHACT} \cite{WPHACT} 
is used for four-fermion final states (see \cite{4fDELPHI} 
for the specific DELPHI implementation). For the \hA~analysis, prepared during an earlier stage of this work, the two-fermion 
and four-fermion SM backgrounds are simulated using the {\tt PYTHIA} \cite{PYTHIA} and {\tt EXCALIBUR} \cite{EXCALIBUR}
generators respectively. Backgrounds from two-photon events are generated using {\tt PYTHIA}.
All background samples have a size equivalent to 20-50 times the luminosity recorded at the corresponding centre-of-mass energy.

The {\tt HZHA} generator is used to generate \hZ~signal samples for 40$\leq$\mh$\leq$120~\massunit,
in steps of 2.5~\massunit. Z boson decays into quarks, neutrinos, electrons and muons are simulated, 
and the Higgs boson is decaying to either a pair of gluons, or a pair of light quarks (s quarks in practice); 
at several mass points, samples with Higgs boson decays into heavy quarks are used to verify the flavour-independence of the results.
For the neutrino channel, signal samples are also generated every~5~\massunit\ for 10$\leq$\mh$\leq$40~\massunit, and 
at \mh$=$4~\massunit; heavy quark samples are simulated at all mass points (when kinematically allowed) in this case. All hZ signal samples contain 5000 events.

The hA process is also simulated using the {\tt HZHA} generator, in the mass range 16$\leq$\mh+\mA$\leq$180~\massunit. 
The simulation starts at \mh,\mA$=$4~\massunit; steps of 5~\massunit\ are used when 10$\leq$\mh,\mA$\leq$30~\massunit; 
10~\massunit\ steps are used otherwise. 
Higgs boson decays into gluon or light quark pairs are simulated over the whole mass range; samples with heavy 
quarks are simulated at several mass points for cross-checks. All hA signal samples contain 2000 events.

For low Higgs boson masses, the decay final state is determined by non-perturbative mechanisms (direct coupling to hadronic resonances).
There is no robust model of this regime, and the efficiencies of the analyses described below are difficult to predict. We therefore 
limit our search to Higgs boson masses above 4~\massunit, where the {\tt PYTHIA} hadronization model is expected to be valid.

\section{Search for hA production}
\label{sec:hAproduction}

The flavour-independent hA search is designed to cover a 
large part of the accessible h and A mass range, and is based on general 
kinematic features such as event shapes and dijet mass information. 

The higher multiplicity of gluon jets compared to quark jets results in a higher selection 
efficiency for hA~\ra~gluons, but also in a worse mass resolution for this signal. To minimize 
biases that may arise from these competing effects, the selection efficiencies are 
determined using samples of hA~\ra~quarks, while the mass reconstruction is done with samples of hA~\ra~gluons.

\subsection{Analysis streams}
\label{sec:hAanalysisstreams}

A first preselection is applied to all events, requiring at least 20 charged 
particles, a total reconstructed energy greater than 60\% of the centre-of-mass energy, and 
an effective centre-of-mass energy after initial state radiation greater than 150 GeV (the method used to estimate the
effective centre-of-mass energy is described in~\cite{sprime}).
The efficiencies of the multiplicity cut are typically 98\% for the \hA~\ra~quarks 
signal samples, and 100\% for the \hA~\ra~gluons samples. 

In the rest of the analysis, three different sets of cuts are designed to
obtain sufficient signal acceptance over a large domain in the (\mh,\mA)-plane. The analysis streams
are described below. 

\subsubsection{Four-jet stream} 

Close to the kinematic limit and when both Higgs bosons have comparable masses, 
a four-jet topology is expected. To analyse this topology, events are clustered 
into four jets with the DURHAM algorithm \cite{durham}. All jets are required to 
have an invariant mass larger than 2 \massunit, and to contain at least two charged 
particles. Events are retained in this stream if their thrust value is below 0.85 and 
if the product of the smallest jet energy and inter-jet angle (called 
$\mathrm{E_{min}\alpha_{min}}$) is greater than 10 GeV$\cdot$rad. Dijet invariant 
mass information is used to reject events compatible  with WW production as in \cite{ZZpaperDelphi}, 
requiring the  corresponding probability, $\mathrm{P_{WW}}$, 
to be less than 0.01. This proves helpful not only when \mh~=~\mA~$\sim$ 80~\massunit, 
but also for other signal masses, where WW production contributes to the expected 
background through wrong jet pairings. 

\subsubsection{Three-jet stream} 

When the h and A mass difference increases, the events tend to contain 
three jets, since the jets resulting from the decay of the lighter Higgs boson cannot always
be resolved. The same behaviour is observed if h and A both have low masses, because 
of the larger boost in this case. To analyse this topology, only events with thrust 
values between 0.70 and 0.92 are kept. Events are then clustered into four jets, 
and events compatible with WW production are rejected as in the four-jet stream. 
The remaining events are clustered into three jets. As before, all jets are required 
to have a mass larger than 2 \massunit\ and to contain at least two charged particles.

\subsubsection{High-thrust stream} 

Finally, when both Higgs bosons are sufficiently light, signal events 
tend to become cigar-like. Events are selected in this analysis stream if they have a 
thrust value larger than 0.92, and are clustered into three jets, each jet having to satisfy the 
same quality criteria as above. As the dominant background comes from two-quark processes, 
the kinematic compatibility with W-pair events is not tested.

\subsubsection{Comparison between data and simulation}
\begin{figure}[htbp]
\begin{center}
  \subfigure{ \includegraphics[width=.4\textwidth]{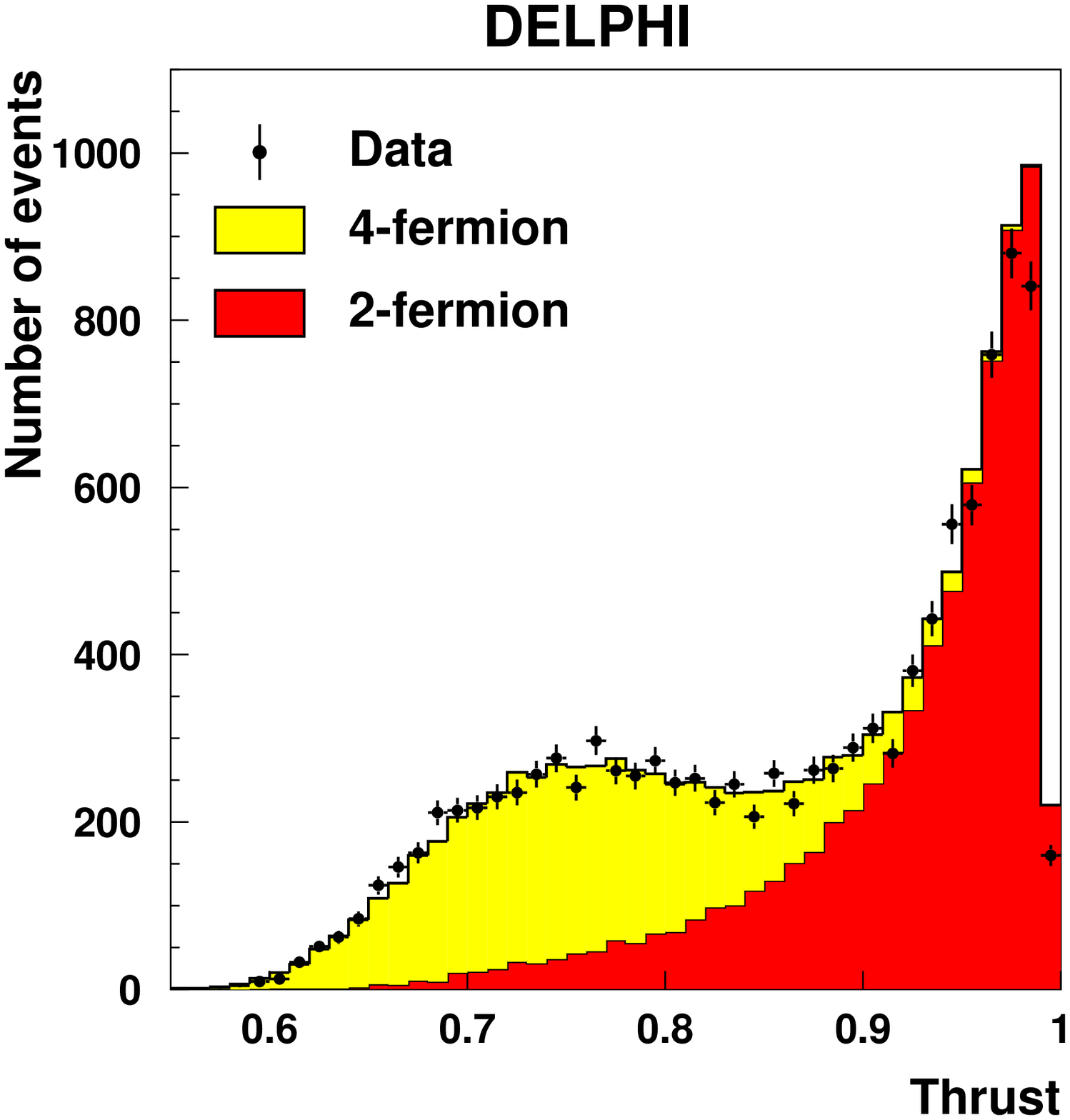}} 
  \subfigure{ \includegraphics[width=.4\textwidth]{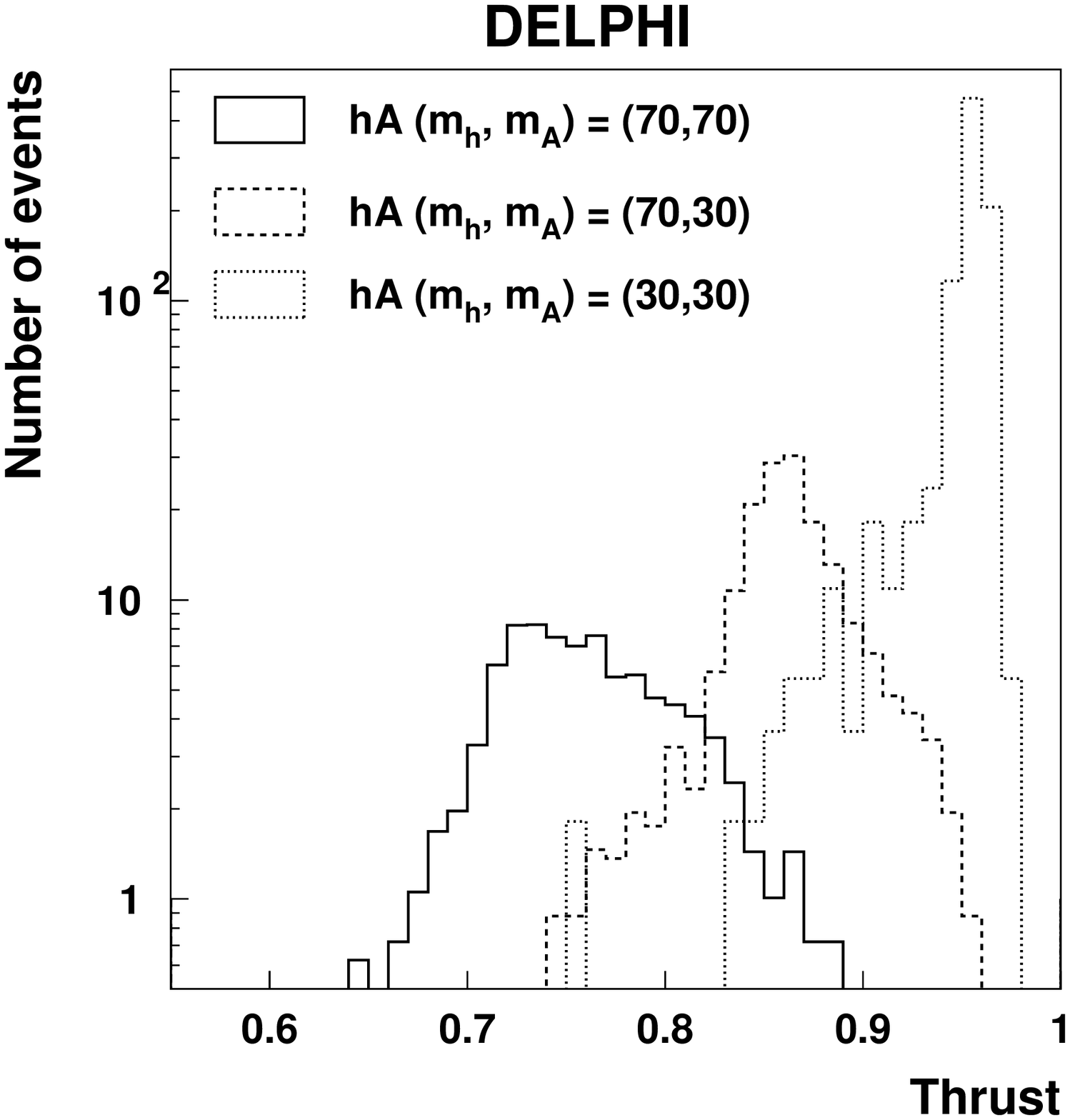}} \\
  \vspace{-0.75cm}
  \subfigure{ \includegraphics[width=.4\textwidth]{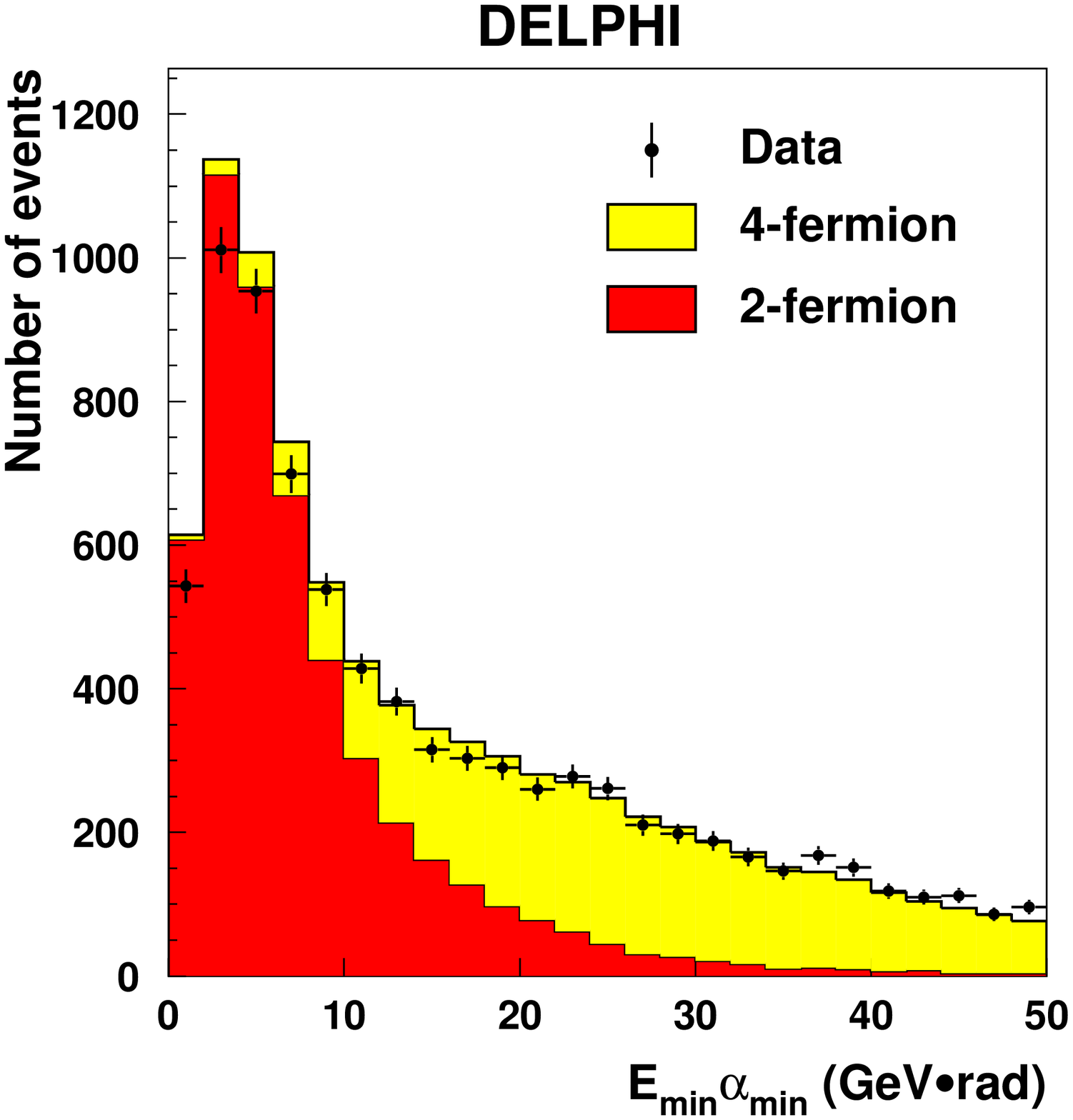}} 
  \subfigure{ \includegraphics[width=.4\textwidth]{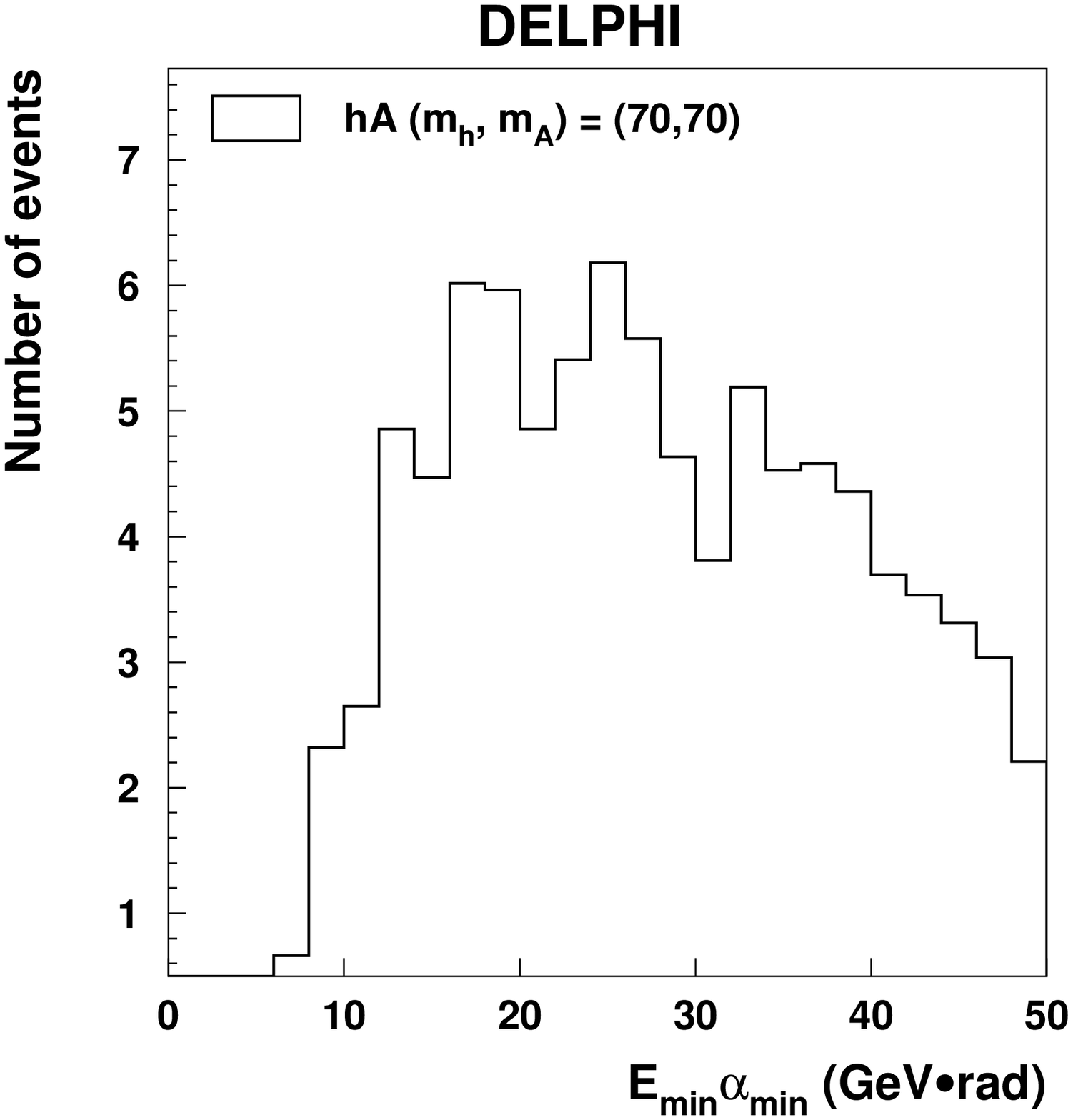}} \\
  \vspace{-0.75cm}
  \subfigure{ \includegraphics[width=.4\textwidth]{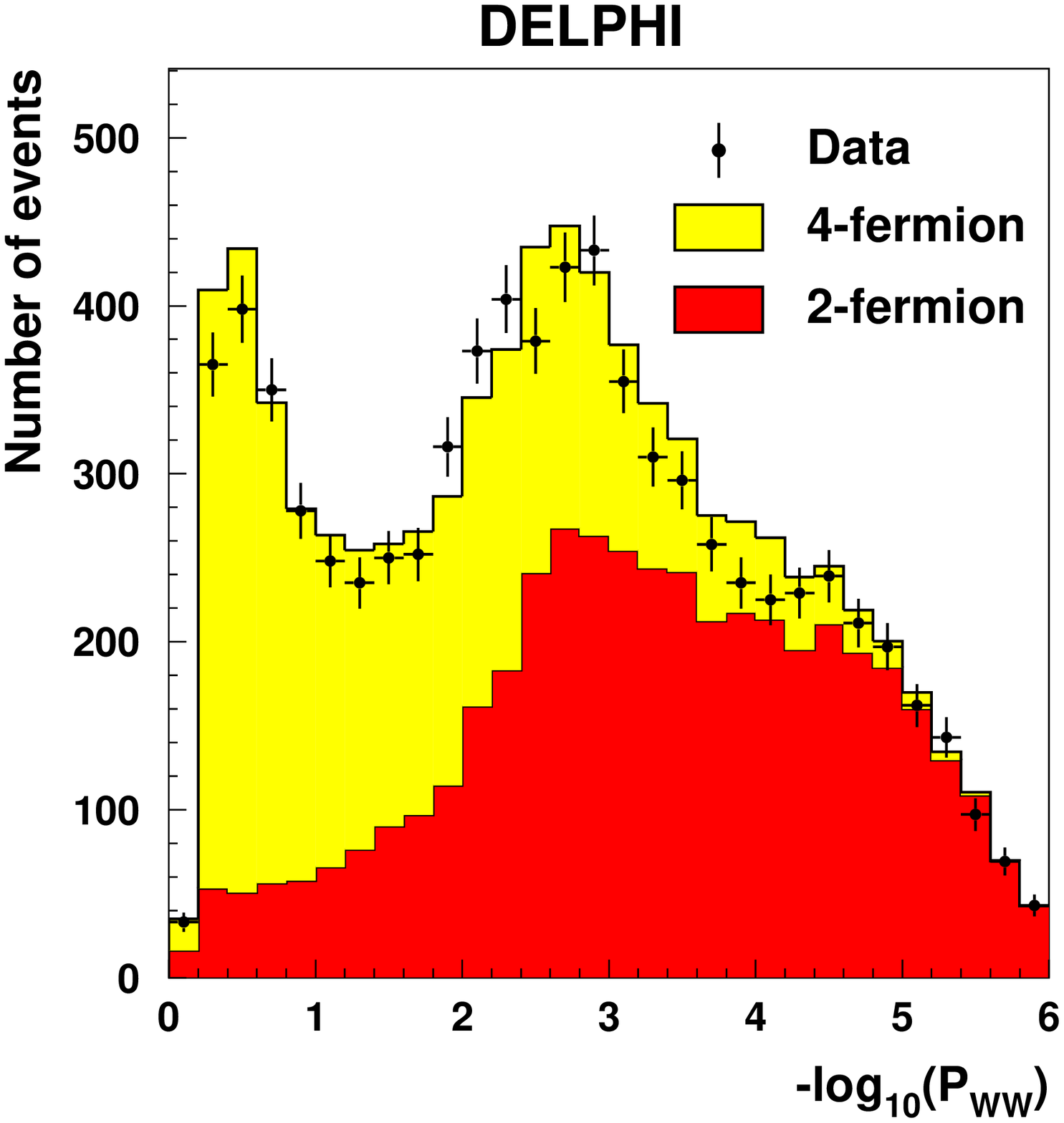}} 
  \subfigure{ \includegraphics[width=.4\textwidth]{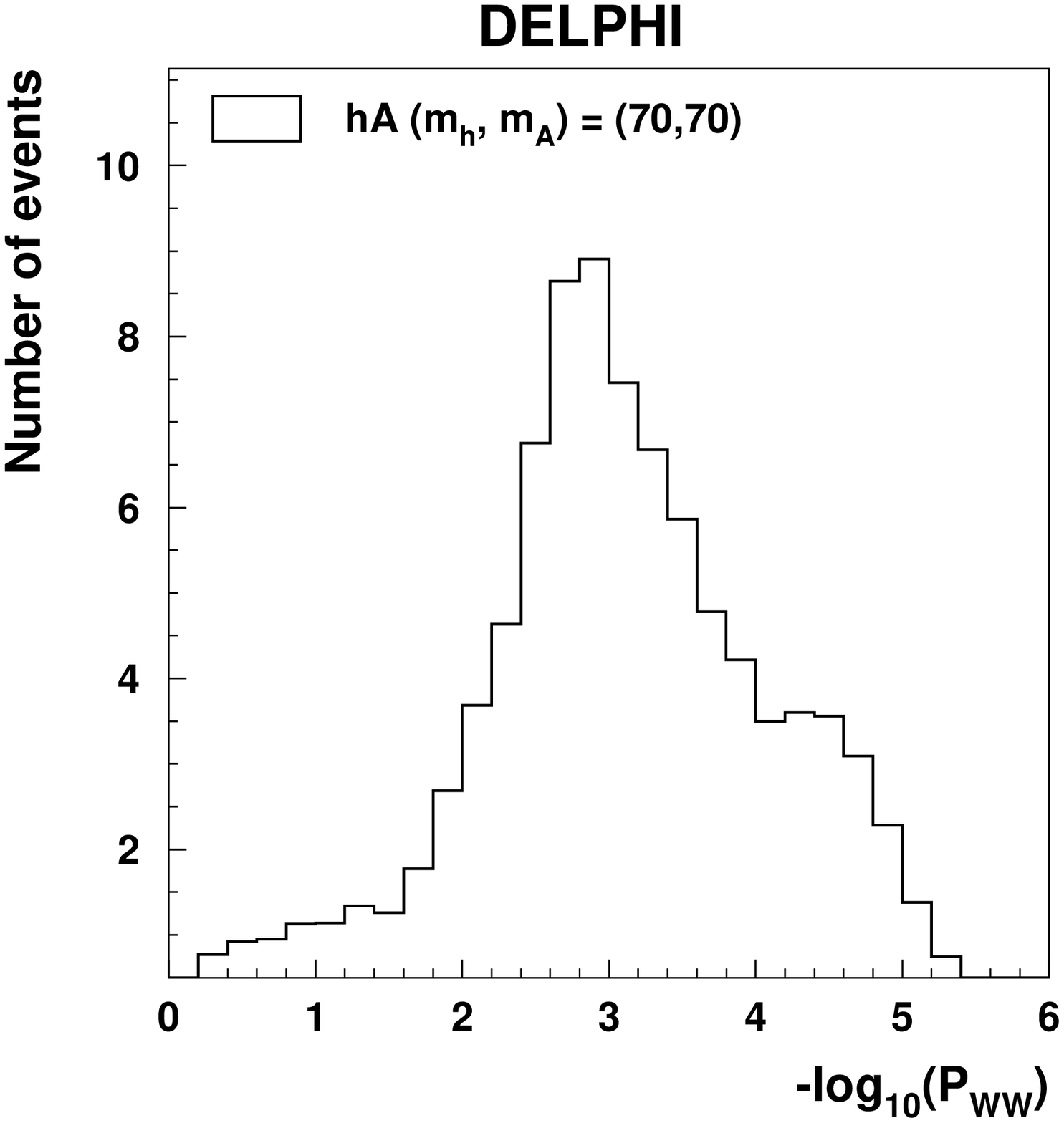}} \\
\end{center}
\vspace{-1em}
\caption{\hA~search. 
         Comparison between data and simulation for the variables 
         used in the three selection streams, at the preselection level. 
         Thrust, $\mathrm{E_{min}\alpha_{min}}$, and -$\mathrm{\log_{10}(P_{WW})}$
         are shown from top to bottom.  The signal normalization assumes \cha=1. 
         The discriminating power is indicated by 
         corresponding distributions from representative signal samples. In particular,
         the thrust distributions justify the use of three complementary selection streams.}
\label{fig:hA.variables}
\end{figure}

Distributions of thrust, $\mathrm{E_{min}}\alpha_{min}$ and $\mathrm{-\log_{10}P_{WW}}$
are shown in Figure \ref{fig:hA.variables} for example signals representative of the three analysis 
streams. Their performance and complementarity are illustrated in Table \ref{tab:hA.efficiencies}. As 
discussed above, these efficiencies are obtained assuming h and A decays into 
light quark pairs. From comparisons with a number of samples with Higgs boson decays into b quarks,
the quark flavour-dependence of the efficiencies is found smaller than 2\%. The efficiencies obtained 
with gluonic Higgs boson decays are higher by a factor up to 1.2 in the three-jet and four-jet analyses, 
and by a factor up to 2 in the high-thrust analysis. A numerical comparison between data and expected background 
is shown in Table \ref{tab:hA.dataMC}. 

\begin{table}[htbp]
 \begin{center}
 \begin{tabular}{l|c|c|c} \hline\hline
               Signal   & $\epsilon_{\mathrm{hA}}^{\mathrm{four-jet}}$(\%)
                        & $\epsilon_{\mathrm{hA}}^{\mathrm{three-jet}}$(\%)
                        & $\epsilon_{\mathrm{hA}}^{\mathrm{high-thrust}}$(\%) \\
   \hline 
   Four-jet events      &   $\sim$ 70           &  $\sim$ 45            &   0-5              \\ 
   Three-jet events     &   $\sim$ 50           &  $\sim$ 70            &   0-5              \\ 
   High-thrust events  &     0-5               &    0-5                &   10-15             \\  \hline
  \end{tabular}
 \end{center}
  \vspace{-1em}
  \caption{hA search. Efficiencies of the three analysis streams on three classes 
           of hA signal events, considered as four-jet events when \mh,\mA~$>$ 60~\massunit; as high-thrust events 
           when \mh,\mA~$<$~30~\massunit; and as three-jet events in the remaining cases.}
 \label{tab:hA.efficiencies}
\end{table}

\begin{table}[htbp]
 \begin{center}
 
  \begin{tabular}{c|c|c|c|c}\hline\hline
   centre-of-mass energy & 189 GeV & 196 GeV & 200 GeV & $>$204.5 GeV \\
   \hline
   Four-jet stream :     &         &         &         &\\
   expected background   & 433.5   & 221.3   & 259.0   & 634.5 \\
   observed events       & 459     & 248     & 232     & 642   \\
   \hline
   Three-jet stream :    &          &         &         &\\
   expected background   & 1593.3   & 750.6   & 797.0   & 1894.3 \\
   observed events       & 1585     & 736     & 772     & 1824   \\
   \hline
   High-thrust stream :  &          &         &         & \\
   expected background   & 1384.9   & 642.5   & 654.3   & 1516.3 \\
   observed events       & 1331     & 612     & 607     & 1450   \\ \hline
   \end{tabular}
   \end{center}
\vspace{-1em}
\caption{hA search. Numerical comparison between data and background simulation,
         after all selections. The year 2000 data are merged in the last column.
         The statistical uncertainty on the background estimates is 2\%.}
 \label{tab:hA.dataMC}
\end{table}                                                                                    

\subsection{Final discriminant variable}
\label{sec:hAfinaldiscriminant}

In all three analysis streams, a four-constraint fit is performed 
on the selected events, requiring total energy and momentum conservation. 
The compatibility of each event with a given mass hypothesis $(\mh,\mA)$
is estimated using the following quantity : 

$$ \chi^2(\mh,\mA) = 
    \left(\frac{ m_{1}^{rec}-\mh }{\delta m_{1}^{rec}}\right)^2
        + \left(\frac{ m_{2}^{rec}-\mA }{\delta m_{2}^{rec}}\right)^2. $$

\noindent In the above expression, $m_{1}^{rec}$ is the mass of a given dijet, $m_{2}^{rec}$ is the mass 
of the opposite dijet in the four-jet stream, and of the opposite jet in the 
other streams, and $\delta m_{1,2}^{rec}$ are the corresponding errors as obtained 
from the kinematic fit. For every event, the jet pairing that minimizes $\chi^2(\mh,\mA)$ is used to compute the discriminant.

The distributions of this variable are computed, for data and background, on a~1~\massunit~$\times$~1~\massunit~grid in the 
$(\mh,\mA)$-plane. 

The signal distributions are first determined at the simulated mass points. 
To construct the signal distributions at mass configurations that are 
not simulated, the histograms are linearly interpolated between the 
three closest simulated points. The validity of this procedure is discussed in
the next section.

\begin{figure}[htbp]
\begin{center}
  \subfigure{ \includegraphics[width=.4\textwidth]{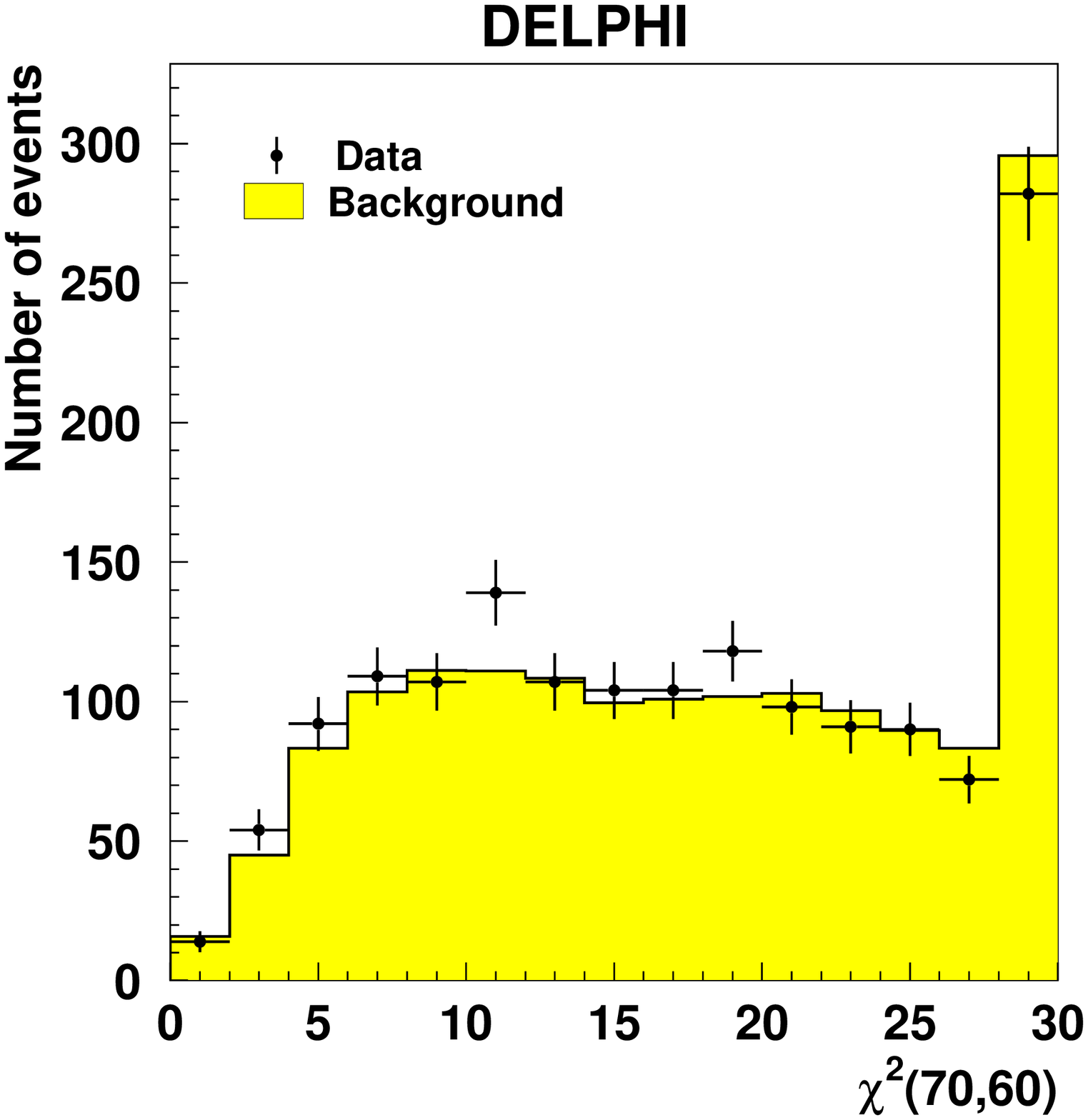}} 
  \subfigure{ \includegraphics[width=.4\textwidth]{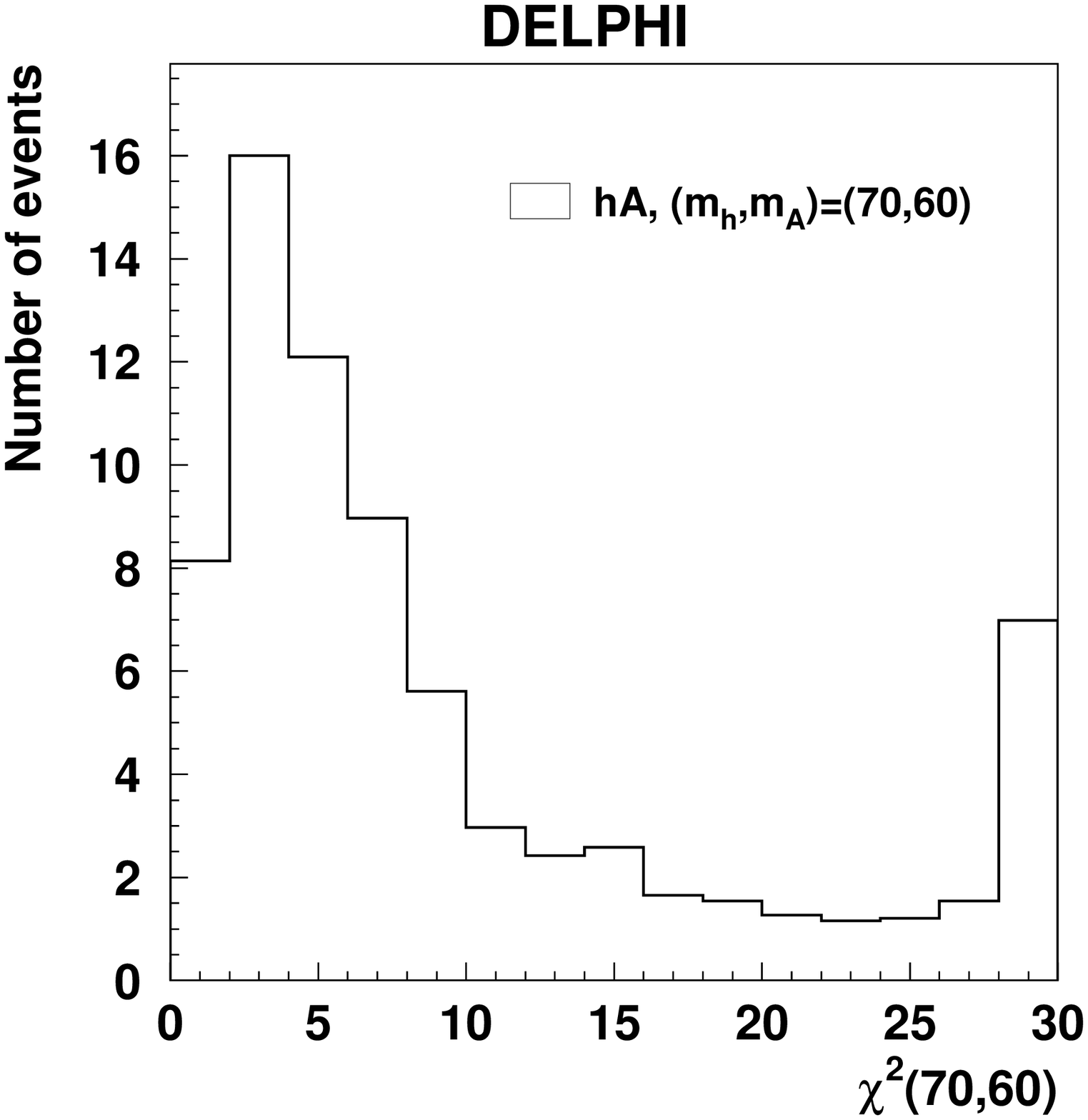}} \\
  \vspace{-0.75cm}
  \subfigure{ \includegraphics[width=.4\textwidth]{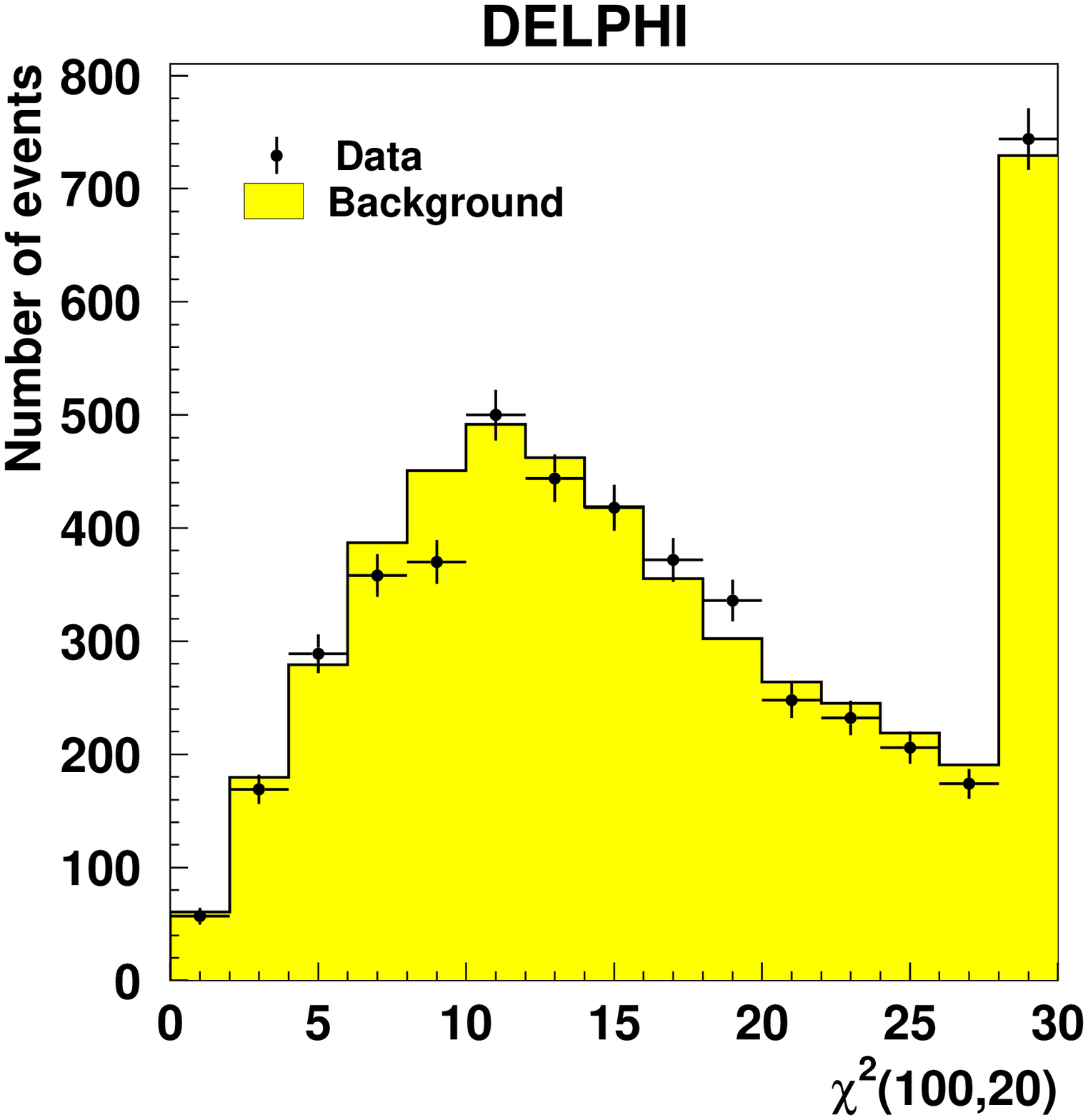}} 
  \subfigure{ \includegraphics[width=.4\textwidth]{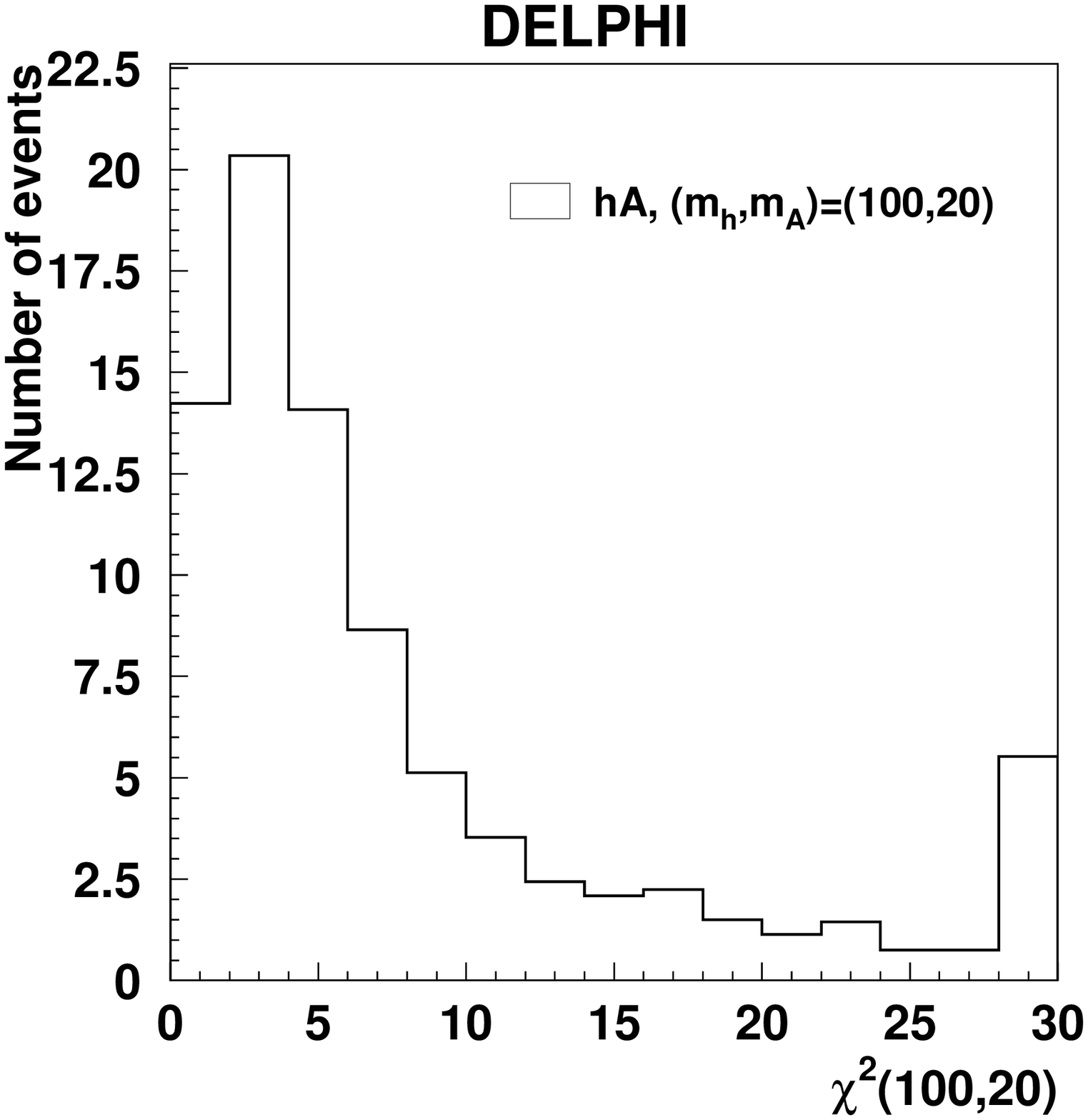}} \\
  \vspace{-0.75cm}
  \subfigure{ \includegraphics[width=.4\textwidth]{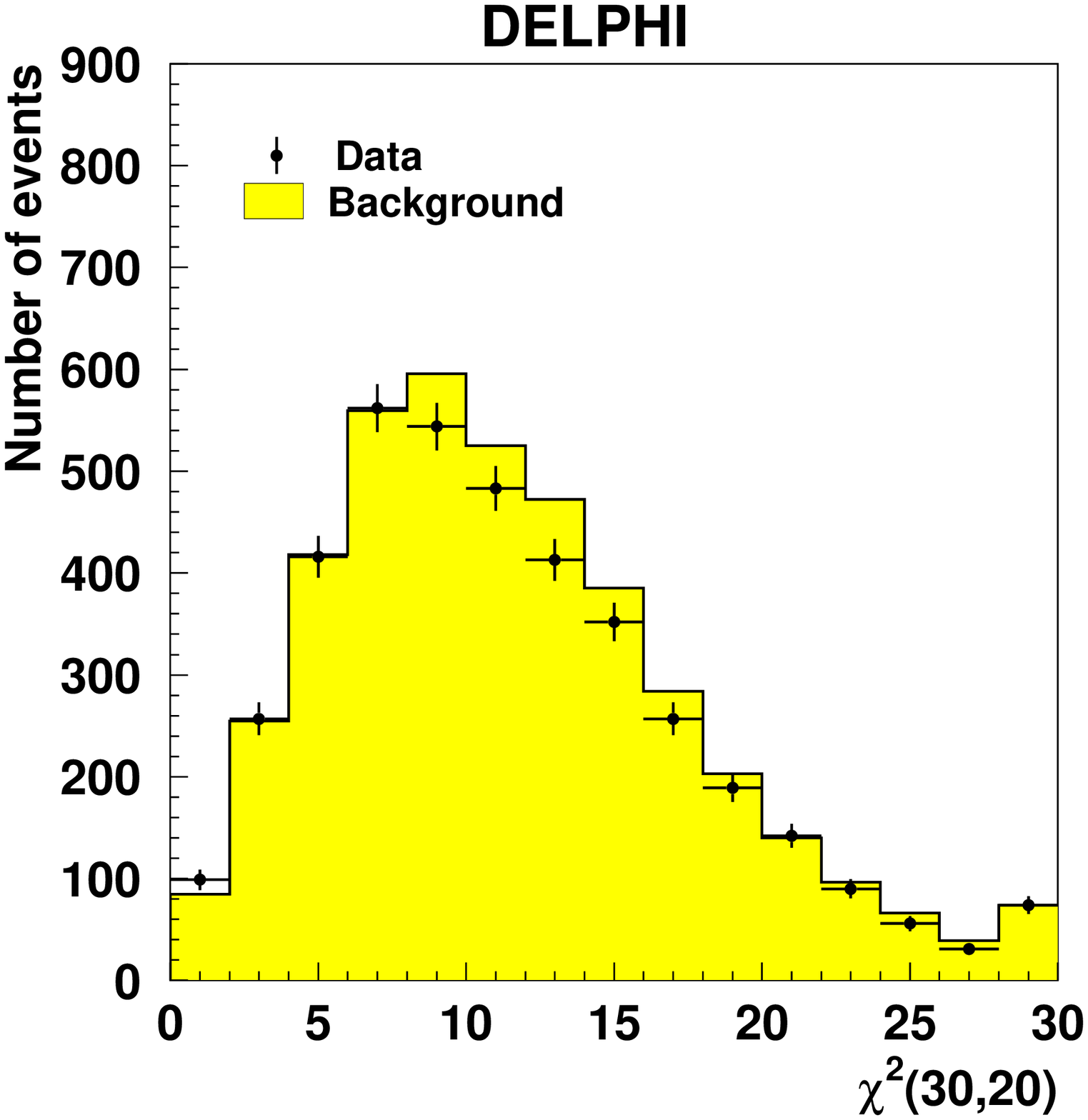}}
  \subfigure{ \includegraphics[width=.4\textwidth]{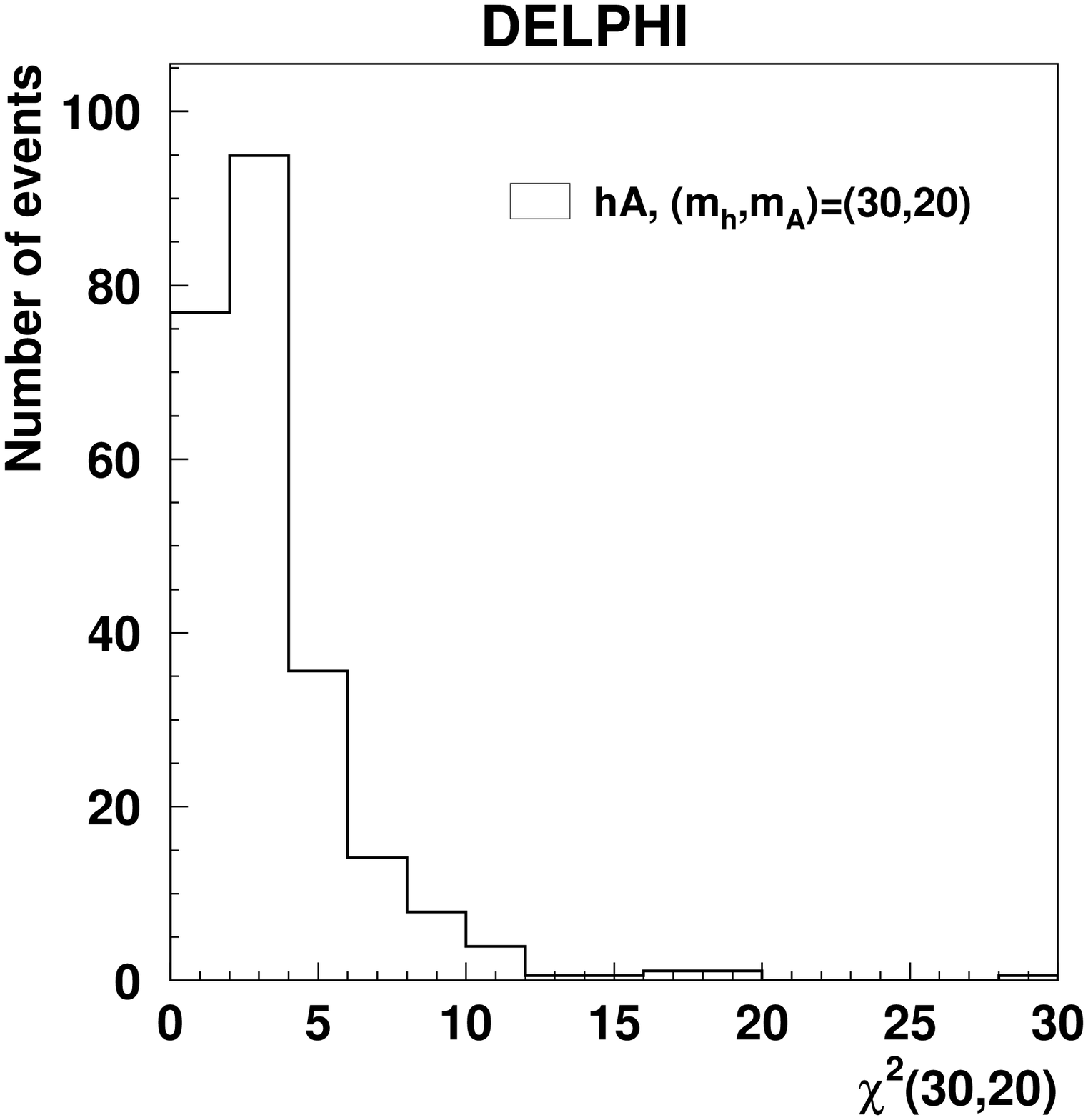}} \\
\end{center}
\vspace{-1em}
\caption{\hA~search. Comparison between data and simulation for the distributions of $\chi^{2}(\mh,\mA)$.
    The discrimination obtained for 
    (\mh,\mA) = (70,60) in the four-jet stream, for (\mh,\mA) = (100,20) in the three-jet stream, and
    for (\mh,\mA) = (30,20) in the high-thrust stream are shown from top to bottom. 
    In all distributions, overflows have been included in the last bin. All the data used for this search are included
    (cf. Section~\ref{sec:dataandsimulation}). The signal normalization assumes \cha=1.}
\label{fig:hA.finaldiscriminant}
\end{figure}

These distributions are used for the statistical evaluation of the compatibility of the data 
with the simulation following the procedure described in Section~\ref{sec:generalstrategy}. 
Figure \ref{fig:hA.finaldiscriminant} illustrates this procedure for three representative example signals.

\subsection{Systematic uncertainties}
\label{sec:hAsystematics}

Systematic uncertainties affecting this analysis are related to imperfect simulation of 
the detector and of SM processes, to residual flavour-dependence of the selections, and to the signal 
interpolation procedure. All effects are discussed below.

The numerical differences at the preselection level between the observed data and the expectation from SM processes
are taken as a first contribution to the uncertainty on the background prediction. For every subsequent selection 
variable, the difference of the average values of the corresponding distributions in data and background is computed; 
the selection cut is then varied by this amount. This procedure leads to an estimated systematic 
uncertainty on the backgrounds of 5\% in the four-jet and three-jet streams, and 10\% in the high-thrust 
stream, at each centre-of-mass energy. The higher uncertainty in the high-thrust stream is mainly due to the observed 
numerical discrepancy after the preselection.

As stated earlier, the flavour-dependence of the selection efficiencies is estimated by 
comparing the results obtained for gluon decays, light quark decays, and b-quark decays. 
The preselection efficiency is slightly worse (by at most 2\%) in the case of light quark decays, 
and is applied to the gluon decays as well. The remaining selections have a quark flavour dependence smaller than
2$\%$. This number is taken as a contribution to the systematic uncertainty on the 
signal rate.

\begin{figure}[htbp]
\begin{center}
  \subfigure{ \includegraphics[width=.4\textwidth]{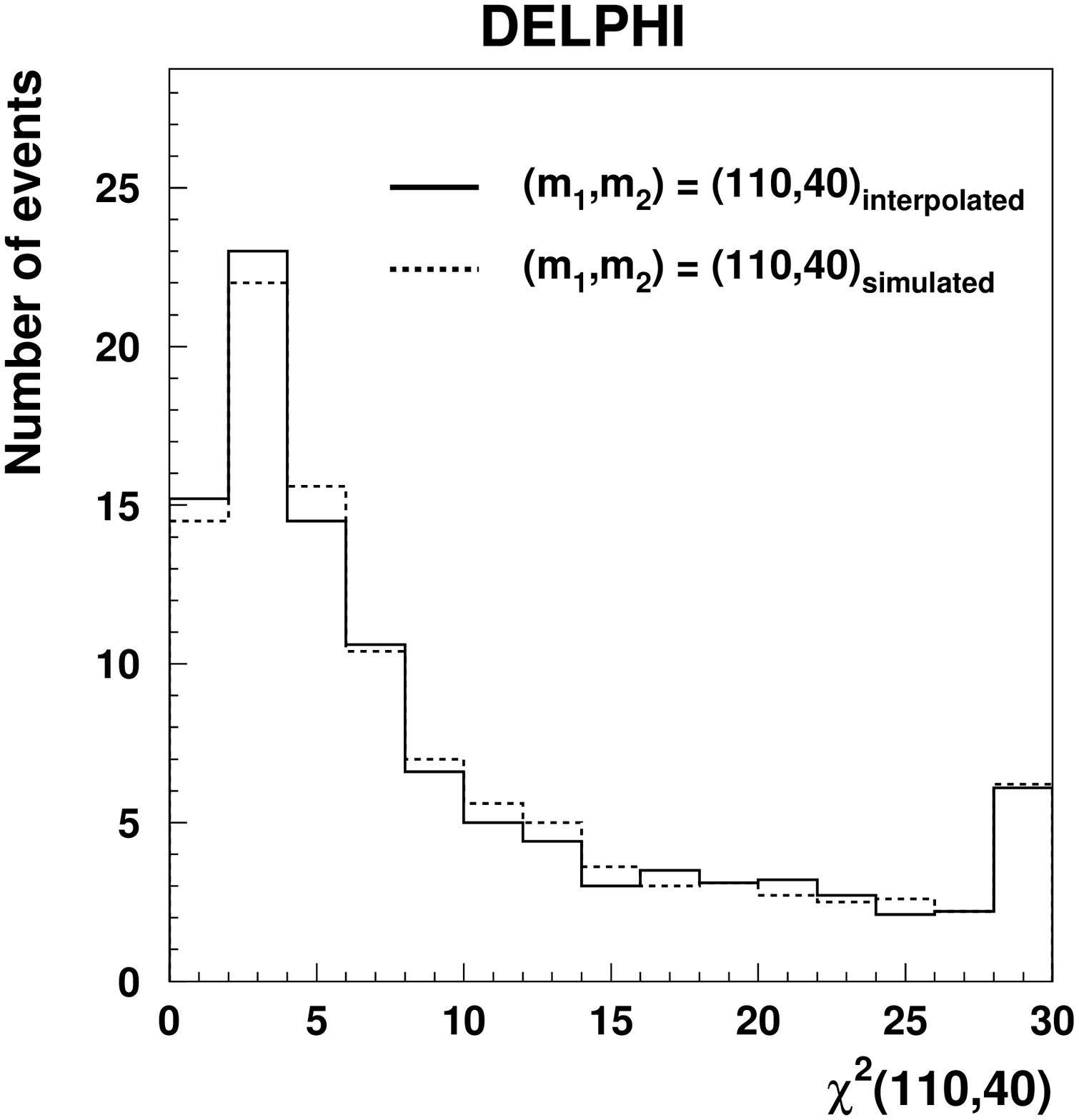}}
  \subfigure{ \includegraphics[width=.4\textwidth]{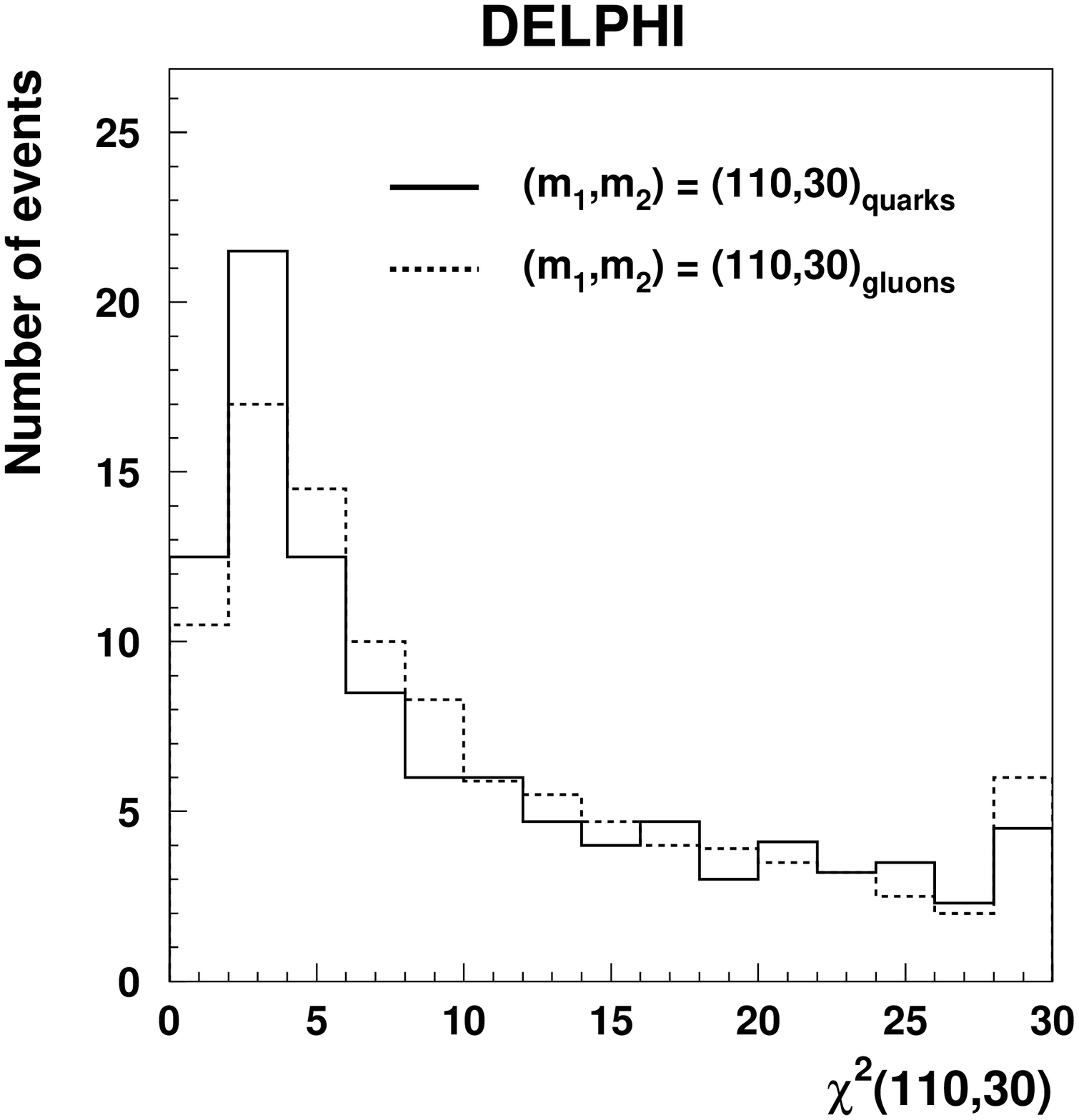}} \\
\end{center}
\vspace{-2em}
\caption{hA search. Left: distribution of $\chi^{2}(110,40)$, as obtained from an interpolation between the nearest simulated 
         samples at (\mh,\mA)=(110,30) and (110,50) \massunit. It is compared with the true (dashed line) 
         distribution from a simulated sample at (\mh,\mA)=(110,40). Right:
         distribution of $\chi^{2}(110,30)$ for Higgs bosons decaying into either 
         \bdqq~(solid line) or \bdgg~(dashed line), reflecting the worse mass resolution obtained for 
         a four-gluon final state. In all distributions, overflows have been included in 
         the last bin.}
\label{fig:hA.interpolation.and.massinformation}
\end{figure}

Possible uncertainties from the interpolation procedure are illustrated in Fig.~\ref{fig:hA.interpolation.and.massinformation},
which compares a simulated distribution with the result of an interpolation between neighbouring points. The 
shapes are well reproduced. To quantify more precisely what accuracy can be expected from this procedure, 
the previous exercise is repeated on a large number of simulated mass configurations. In 
each case, the bin contents of the simulated and interpolated histograms are compared. The 
distribution of the ratio of true and interpolated bin contents has a spread of 5\%, which is 
taken as a second contribution to the systematic uncertainty on the signal rate. Note that 
this is a conservative estimate as the real interpolation is over smaller intervals than the
20 \massunit\ used here.

Figure \ref{fig:hA.interpolation.and.massinformation} also shows the discriminant variable distribution
obtained for quarks and gluons, at (\mh,\mA)=(110,30). 
As expected, the worse mass resolution on gluon jets translates into a somewhat reduced discriminating power; this is 
true for all mass points. The present analysis, using a selection efficiency determined on light-quark signal 
samples and a kinematic analysis calibrated on gluon samples, thus has no bias towards a given flavour 
that would invalidate the results when considering other Higgs boson hadronic decays.


\section{Search for hZ production}
\label{sec:hZproduction}

This section describes a flavour-independent search for the \epem\ra~\hZ\ process. The analyses used in this section are adapted 
from those used for the DELPHI \epem\ra~\ZZ\ and \epem\ra~\Zgstar\ cross-section 
measurements~\cite{ZZpaperDelphi,ZgpaperDelphi}, or from the previous section. 
The analyses are briefly described in the following; we refer to the original publications for further details.

\subsection{Final states with jets}
\label{sec:hZqqqq}

The hZ search in the fully hadronic final state is described below.
Two mass regions are analysed separately, namely \mh$<$40~\massunit\ using a three-jet analysis, 
and \mh$>$40~\massunit\ using a four-jet analysis.

\subsubsection{Low mass search (\mbox{\boldmath $\mh < 40$ \massunit})}

The region 4~\massunit$<$\mh$<$40~\massunit\ is analysed using the hA three-jet stream, described 
in section \ref{sec:hAanalysisstreams}. Efficiencies and discriminant variable 
distributions are determined from interpolations between existing hA signals, obtaining distributions of
$\chi^2(\mh,\mZ)$, where \mZ\ is the Z boson mass and \mh\ is ranging from 4 to 40 \massunit.

A few \hZ~samples have been simulated (at \mh~=4, 20 and 40 \massunit) to verify the validity of this procedure. 
Despite the different Higgs boson angular distributions and the Z boson width, the efficiencies and 
$\chi^{2}(\mh,\mZ)$ distributions proved accurate within the systematic errors estimated in the hA three-jet channel.

\subsubsection{High mass search (\mbox{\boldmath $\mh > 40$ \massunit})}

The selection used to analyze the four-jet channel in the \epem\ra~\ZZ~cross-section
measurement \cite{ZZpaperDelphi} is adapted to test the hypothesis of Higgs boson production. 
After the selection of high multiplicity events with at least four DURHAM jets,
the only remaining backgrounds are from ZZ and WW four-quark final states, and four jet events arising 
from \bdqq~production with hard final state gluon radiation. Event shape variables and jet kinematics are 
used to further enhance the signal. 

A constrained fit is performed on the reconstructed jets, requiring energy and momentum conservation
between the initial and final state. The fitted jet momenta are then used to reconstruct the dijet masses for all
possible jet pairings. The uncertainties on the fitted momenta and their correlation matrix are 
taken into account in that procedure. The dijet masses and their uncertainties are used to compute the probability 
that each event is compatible with WW, ZZ, \bdqq, or hZ production (in this case, the event probability depends on the signal
mass hypothesis), as in \cite{ZZpaperDelphi}. 

Finally, for each jet, the probability that it originates from a b quark is computed, using the methods 
described in~\cite{btag}. Adding this information in the characterization of Z boson dijets allows better sensitivity 
to the fraction of the signal where the Z boson decays into b quarks. 
This information is not used on the Higgs boson side, thus preserving flavour-independence. 

The method employed to combine the various mass, event shape and b-tagging probabilities using likelihood ratio 
products is described in \cite{ZZpaperDelphi}. The resulting discriminant variable is called ${\mathrm P_{Higgs}}$.

\begin{figure}[htbp]
\begin{center}
\subfigure{ \includegraphics[width=.4\textwidth]{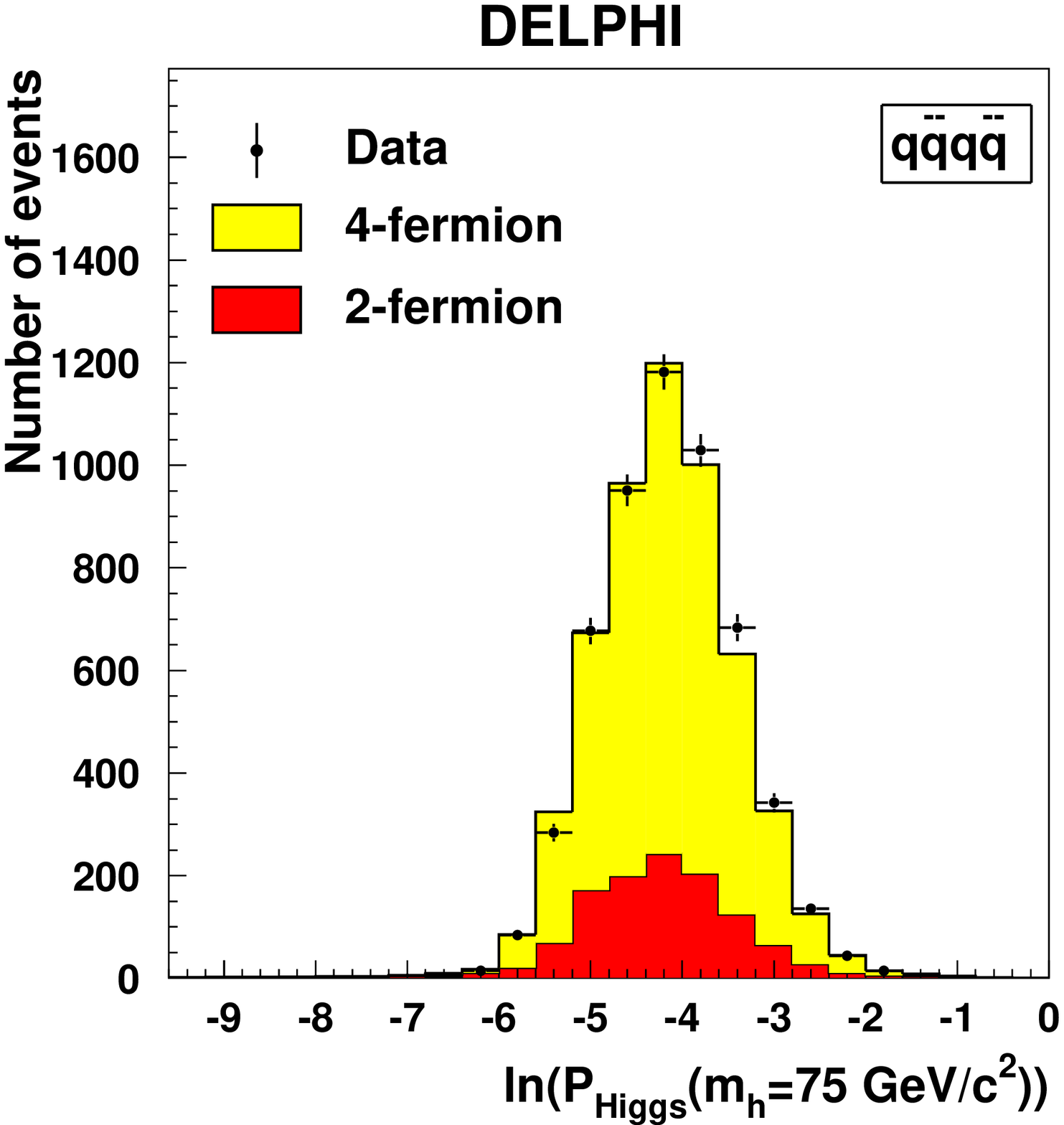}} 
\subfigure{ \includegraphics[width=.4\textwidth]{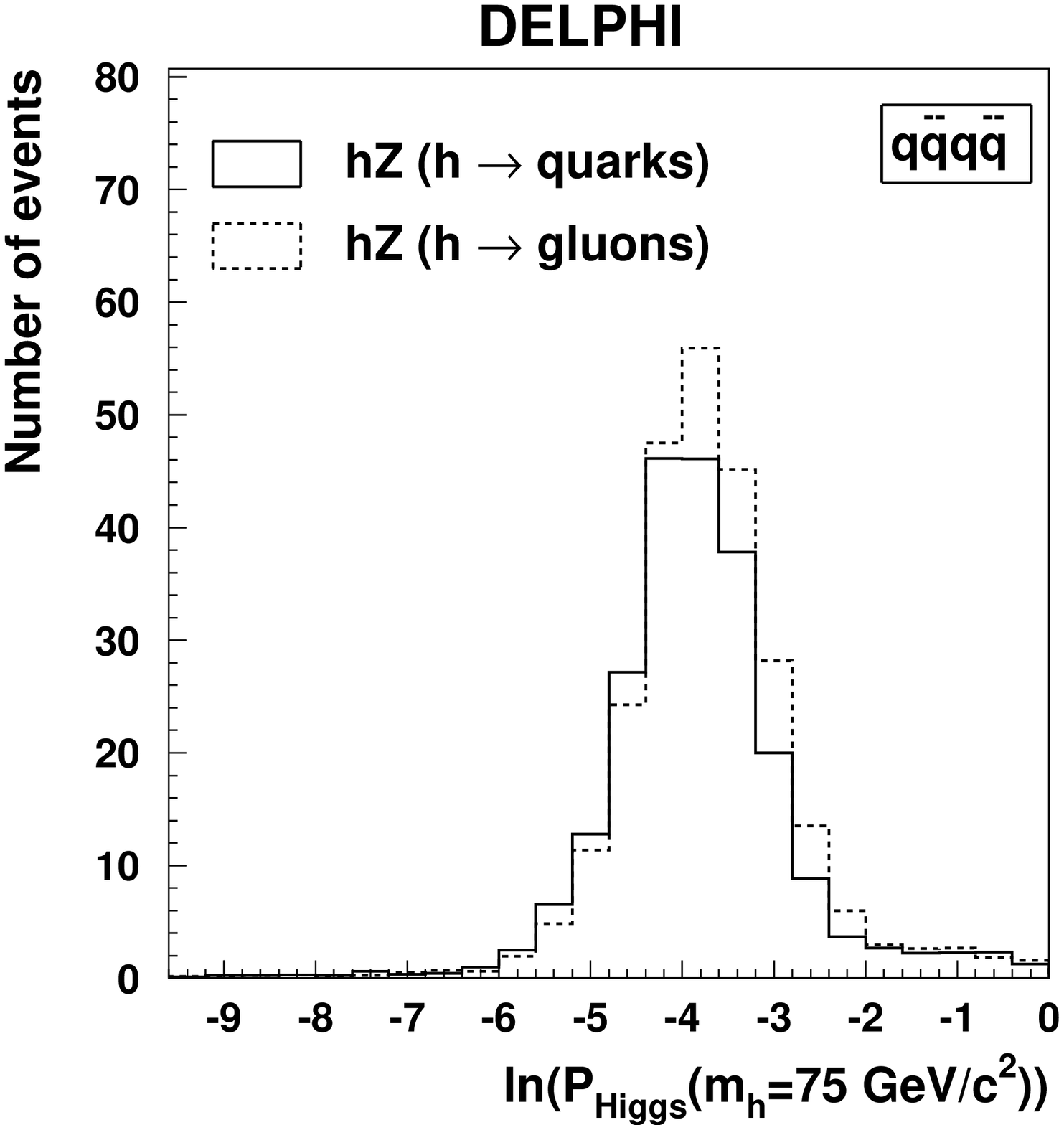}} \\
\vspace{-0.75cm}
\subfigure{ \includegraphics[width=.4\textwidth]{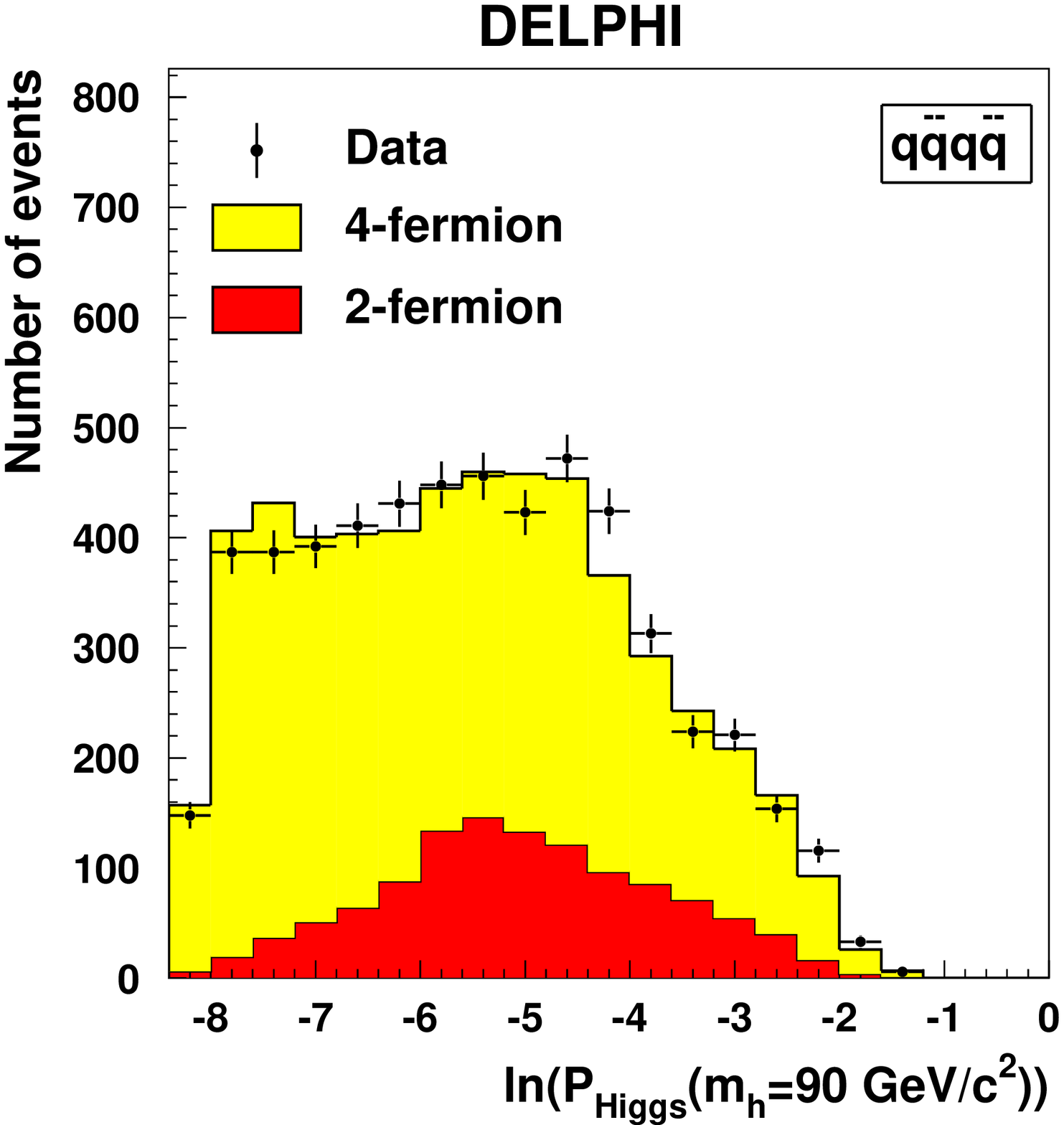}} 
\subfigure{ \includegraphics[width=.4\textwidth]{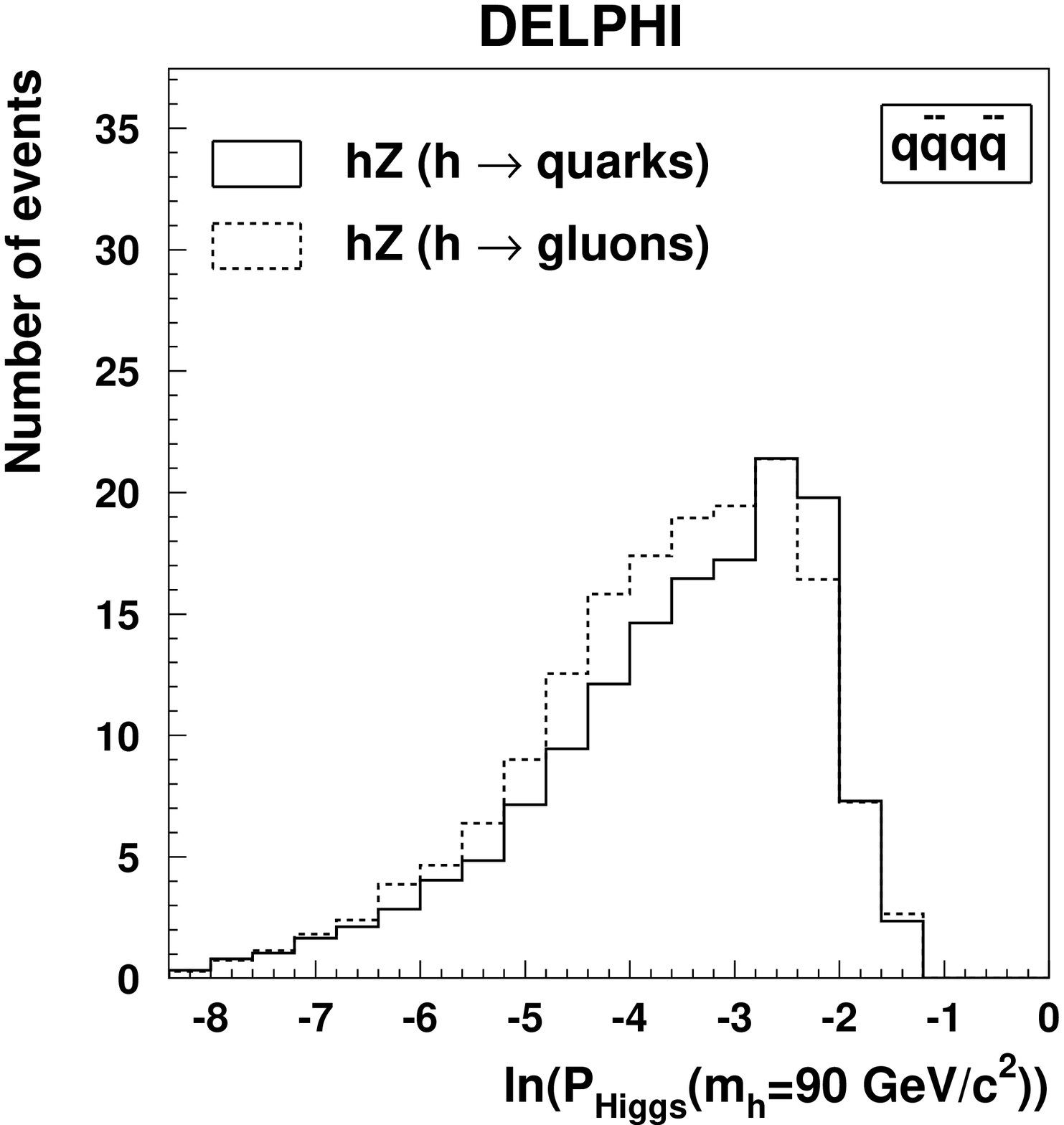}} \\
\vspace{-0.75cm}
\subfigure{ \includegraphics[width=.4\textwidth]{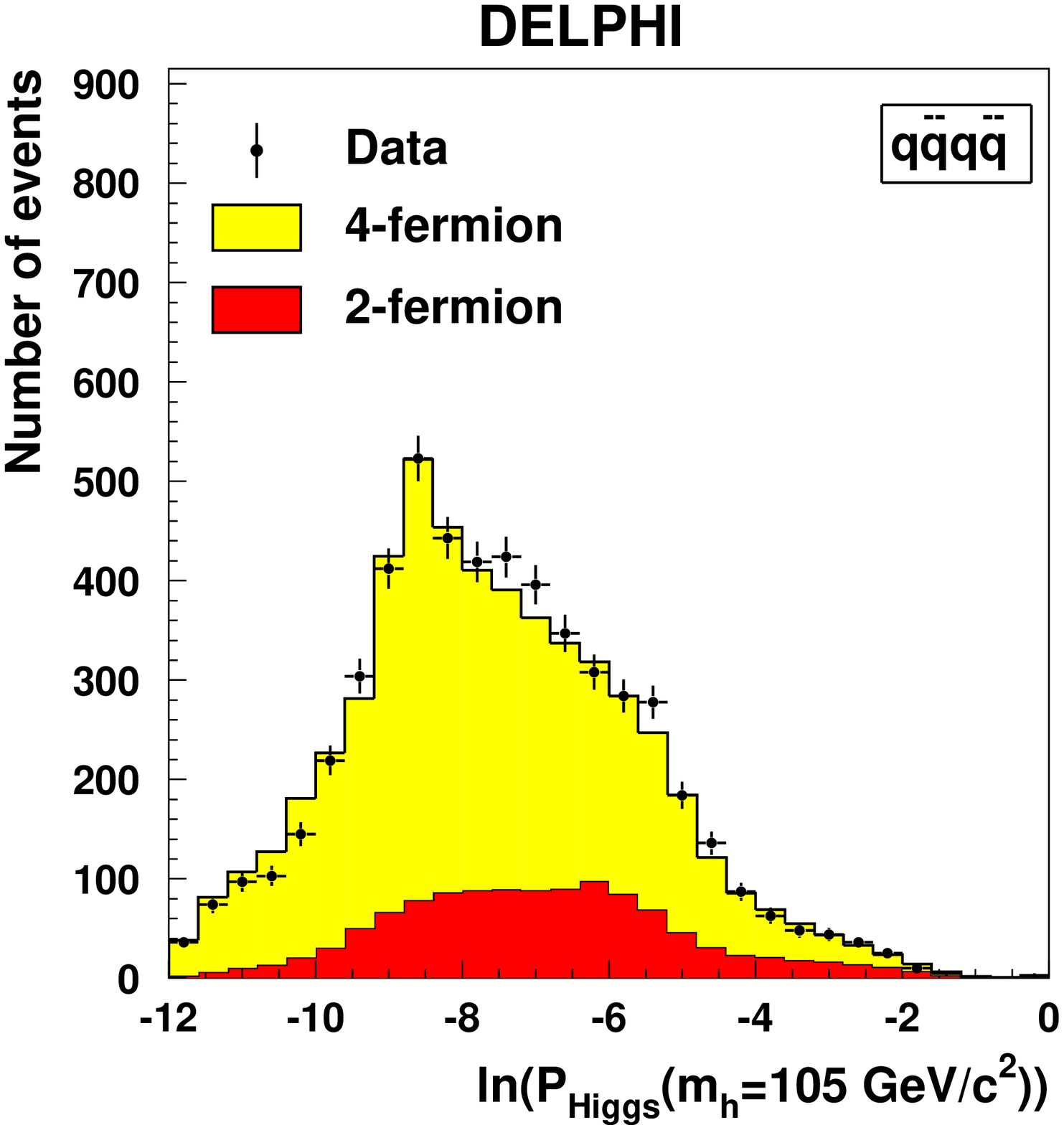}} 
\subfigure{ \includegraphics[width=.4\textwidth]{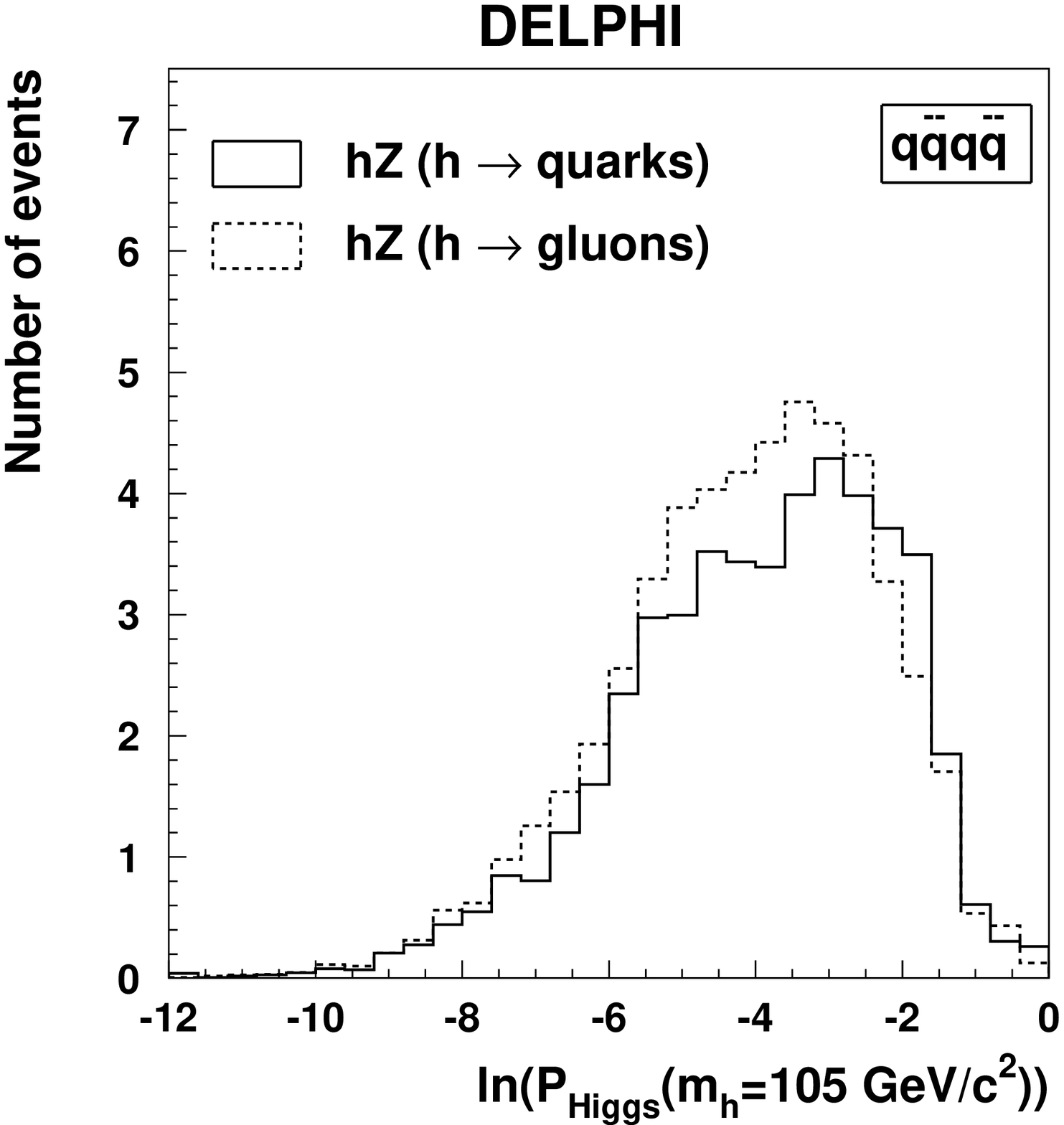}} \\
\vspace{-0.75cm}
\end{center}
\caption{hZ search, final state with jets.
Comparison between data and simulation for the distributions of ln(P$\mathrm{_{Higgs}}$). 
Higgs boson mass hypotheses of 75, 90 and 105~\massunit\ 
are shown from top to bottom. All data from 189 to 209 GeV are included. The signal normalization assumes \chz=1.}
\label{fig:hZqqqq}
\end{figure}

Table~\ref{tab:hZ.4jets.DataMC} gives a numerical comparison between data and simulation, for a 
few Higgs boson masses. Examples of ln(P$\mathrm{_{Higgs}}$) distributions are shown in Figure~\ref{fig:hZqqqq}. 
At each Higgs boson mass hypothesis, the P$\mathrm{_{Higgs}}$ distributions in data, background and signal 
are used for the statistical interpretation of the results (see Section~\ref{sec:generalstrategy}). 
For every mass hypothesis, the signal distribution is obtained from an interpolation between the two closest simulated
signal samples.

The systematic uncertainties are dominated by the uncertainty on the four-jet 
rate in \bdqq\ events \cite{QCDWorkshop}, conservatively taken to be 10\% 
over the full Higgs boson mass range. Other contributions, evaluated in detail in 
\cite{ZZpaperDelphi}, are small and can be neglected.

\begin{table}[htbp]
  \begin{center}
    \begin{tabular}{c | c c | c r }
      \hline
      \hline
      \mh~(\massunit)  & SM (no \hZ) &  observed & {\large $\epsilon_{\mathrm{hZ}}$}(\%)  & \hZ(\h \ra \bdqq)\\
      \hline
      ~40                 &  ~635.0    & ~659~~~~~      &      31.3                 &  154.6 \\
      ~50                 &  ~522.8    & ~532~~~~~      &      30.1                 &  132.3 \\
      ~60                 &  1784.1    & 1824~~~~~      &      51.9                 &  197.1 \\
      ~70                 &  5457.3    & 5476~~~~~      &      78.6                 &  249.3 \\
      ~80                 &  1681.7    & 1764~~~~~      &      54.2                 &  136.0 \\
      ~90                 &  ~957.4    & ~970~~~~~      &      54.6                 &  ~96.2 \\
      100                 &  ~368.5    & ~372~~~~~      &      47.0                 &  ~41.1 \\
      110                 &  ~~87.6    & ~~75~~~~~      &      33.9                 &  ~~9.4 \\
     \hline
     \end{tabular}
  \end{center}
  \caption{hZ search, final state with jets. 
           Numerical comparison between background simulation, data, and a few example signals, 
           after an illustrative cut on P$\mathrm{_{Higgs}}$ that maximises the 
           product of signal efficiency and purity. 
           All data from 189 to 209 GeV are included. The signal normalization assumes \chz=1. 
           The statistical uncertainty on the signal efficiency and on the background is less than 1\%.}
  \label{tab:hZ.4jets.DataMC}
\end{table}

\subsection{Final states with jets and missing energy}
\label{sec:missingenergy}

The event topology in the missing energy channel changes significantly as a function of the 
Higgs boson mass. For low enough Higgs boson masses a mono-jet final state is expected, while 
at high mass the jets are well separated, resulting in a two 
jet final state. For both mass domains (with a transition set at \mh=40~\massunit), dedicated analyses 
are described below.

\subsubsection{Low mass search ({\boldmath $\mh < 40$ \massunit})}
\label{sec:missingenergy.monojet}

The analysis used to measure the  \Zgstar~\ra~\qqnunu~cross-sections \cite{ZgpaperDelphi}  
is adopted without major modifications to test the hypothesis of low mass 
Higgs boson production in the missing energy channel. After a preselection reducing the \gammagamma, Bhabha and \Zg\ backgrounds, 
three selections are designed, focusing on different expected topologies as a function of the mass of 
the Higgs boson.

The first selection is optimized to probe very low Higgs boson masses, using an explicit 
cut on the visible mass of the event, which is required to be below 6~\massunit. The second 
selection exploits the large energy imbalance of \hnunu~final states: events are split 
into two hemispheres according to the plane perpendicular to the direction of the thrust axis,
and are required to have one of the hemispheres containing at least 99\% of the total visible 
energy in the event. The third selection, which is less efficient at very low masses because 
of an explicit cut on the charged particle multiplicity, is mainly based on a topological requirement: 
events are forced into a two jet configuration and an upper cut on the opening angle of the two 
jets is set at 78 degrees. All three selections use the information from the veto counters, 
by rejecting events with hits in veto counters far away from energy depositions in calorimeters 
or reconstructed tracks. The three analyses are combined on an event-by-event basis, by selecting 
events that pass at least one cut.

\begin{figure}[htbp]
\begin{center}
   \subfigure{ \includegraphics[width=.4\textwidth]{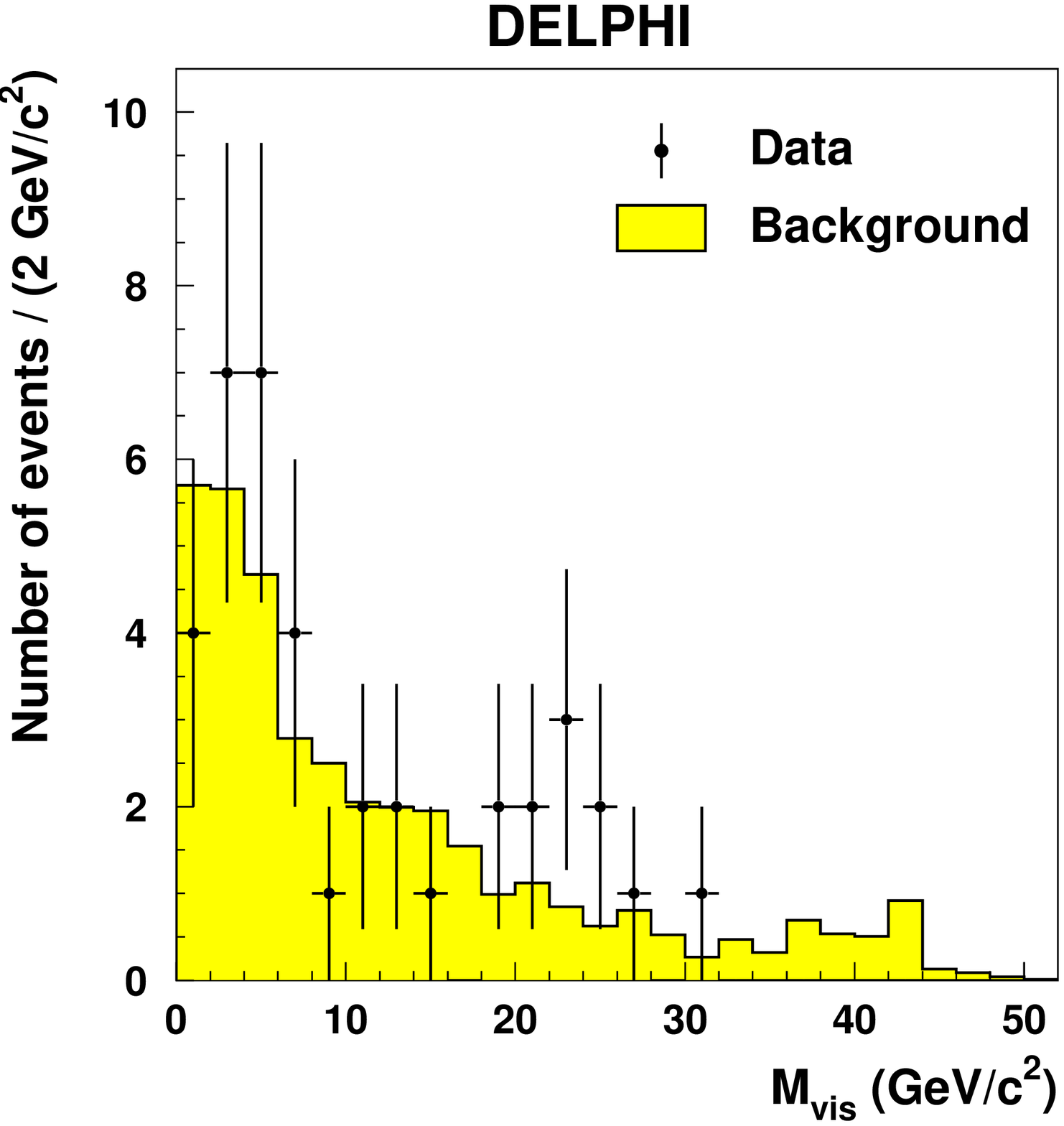}} 
   \subfigure{ \includegraphics[width=.4\textwidth]{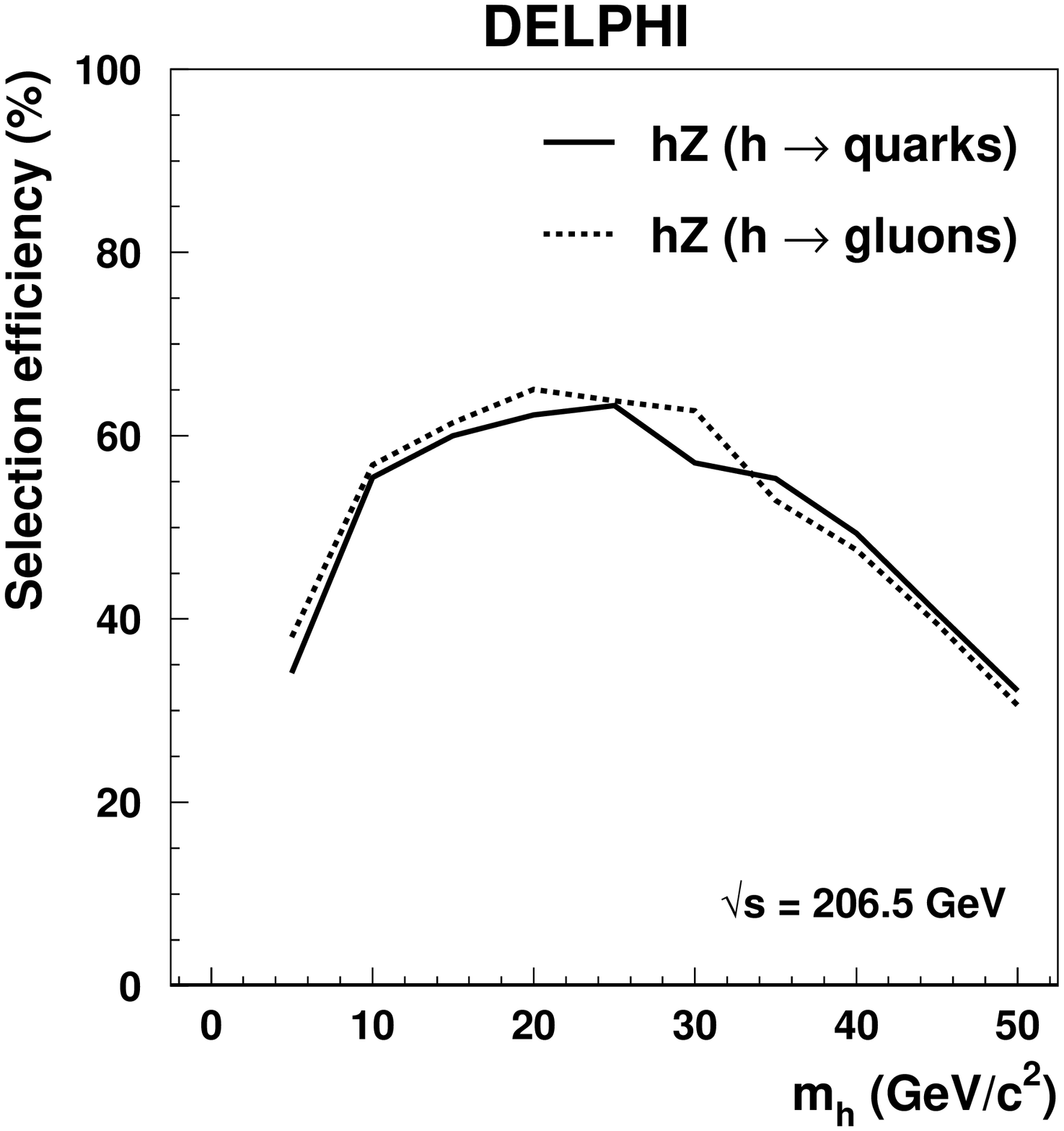}} 
\end{center}
\vspace{-2em}
\caption{hZ search, final state with jets and missing energy, low mass search.
Left:~comparison between data and simulation for the hadronic invariant mass distribution,
after the mono-jet \qqnunu~event selection. All data from 189 to 209 GeV are included. Right:~selection efficiency as a function 
of the Higgs boson mass, at $\sqrt{s}=206.5$ GeV.}
\label{fig:hz.hvvlowmass}
\end{figure}

After all selections, the most important remaining background processes are \Zgstar~and \Wenu.
Higgs boson signal efficiencies reach up to 65\%, and drop to 40\% for masses below 5~\massunit\ and above 40~\massunit.
The efficiencies are found to be independent, within their statistical uncertainties, of the quark or gluon nature of 
the Higgs boson decay.
The reconstructed visible mass distribution is used as discriminant variable for the statistical 
evaluation of the compatibility of the data with the expectation for different Higgs boson mass hypotheses 
(see Section~\ref{sec:generalstrategy}).
For every mass hypothesis, the signal distribution is obtained from an interpolation between the two closest simulated
signal samples. Higgs boson decays to gluons are assumed, given the worse mass resolution expected in this case.
The reconstructed mass spectra and Higgs boson signal efficiencies are illustrated 
in Figure~\ref{fig:hz.hvvlowmass}.

The systematic uncertainty on the predicted background rate is 9\%, obtained 
by varying each cut by an amount equal to the difference of the corresponding variable's average values 
in data and in simulation, at the preselection level.
The uncertainty also includes a 5\% contribution from the statistical uncertainty on the simulated 
background sample \cite{ZgpaperDelphi}). The uncertainty on the signal selection efficiency is 5\%.

\subsubsection{High mass search ({\boldmath $\mh > 40$ \massunit})}
\label{sec:missingenergy.twojets}

The selection described below closely follows the search for invisibly decaying 
Higgs bosons produced together with \Z~bosons decaying into hadrons~\cite{DELPHIinvhiggs}.
Backgrounds originating from \qcd~and \gammagamma~processes are rejected by applying the same selections as 
in~\cite{DELPHIinvhiggs}. Events with more than two identified leptons are then rejected. After this step, all events 
are clustered into both two-jet and three-jet configurations, using the DURHAM clustering algorithm. 

To obtain a good performance over the mass range 40$<$\mh$<$115~\massunit, three mass 
windows are defined (40$\le$\mh$<$70~\massunit, 70$\le$\mh$<$90~\massunit, and 90$\le$\mh$<$115~\massunit, 
subsequently referred to as first, second, and third mass range, respectively), 
and each is treated with a dedicated analysis. 
The discrimination is obtained in each case using a two-step Iterative Discriminant 
Analysis (IDA) \cite{IDA}. After the first iteration, a cut on the IDA output 
is made to remove part of the background while retaining 90\% of the signal. The IDA is 
then trained again on the events passing this selection. At each step, the IDA is trained simultaneously on 
\bdss, \bdcc~and \bdbb~flavours, resulting in a comparable performance for all flavours. 


Twelve variables compose the IDA discriminant for the first mass range. These are the scalar and vector sums of 
the transverse momenta of all reconstructed particles, the total visible energy, the event thrust, the energy of the 
least energetic jet in the three-jet configuration, the difference between the Fox-Wolfram moments $\mathrm{H2}$ and 
$\mathrm{H4}$ \cite{FoxWolfram}, the energy of the most isolated particle, the complement of the two jets opening angle 
(acollinearity), the angle between the two planes formed by the jet axes and the beam axis (acoplanarity), 
the total missing mass, and the highest transverse momentum of a particle with respect to its jet in the two-jet 
configuration. The IDA was trained with signal samples ranging from \mh=40 to 67.5~\massunit.


The analysis in the second mass range uses the same twelve variables as in the low mass 
analysis, except for the energy of the least energetic jet in the three-jet configuration, which 
is replaced by the event b-tag probability as defined in \cite{btag}. The IDA was trained with signal samples ranging 
from \mh=70 to 87.5~\massunit.


In the third mass range, the IDA is composed of the same variables as above, but 
cuts removing the tails in the distributions of the input variables are applied prior to the training,
as is done in~\cite{DELPHIinvhiggs}. For every centre-of-mass energy, the IDA was trained on signal samples 
ranging from \mh=90~\massunit\ to the kinematic limit.

Numerical comparisons between data and simulation are given in Table~\ref{tab:hZ.qqnunuhighmass.DataMC}, after 
the cut on the first IDA output.

The measured jet momenta are rescaled so that the missing mass equals \mZ, and the hadronic mass is computed 
using the rescaled jets. For each signal hypothesis, a window around the nominal 
Higgs boson mass is defined as a final selection, to maximize the product of signal efficiency and purity. 

\begin{figure}[htbp]
\begin{center}
\subfigure{ \includegraphics[width=.4\textwidth]{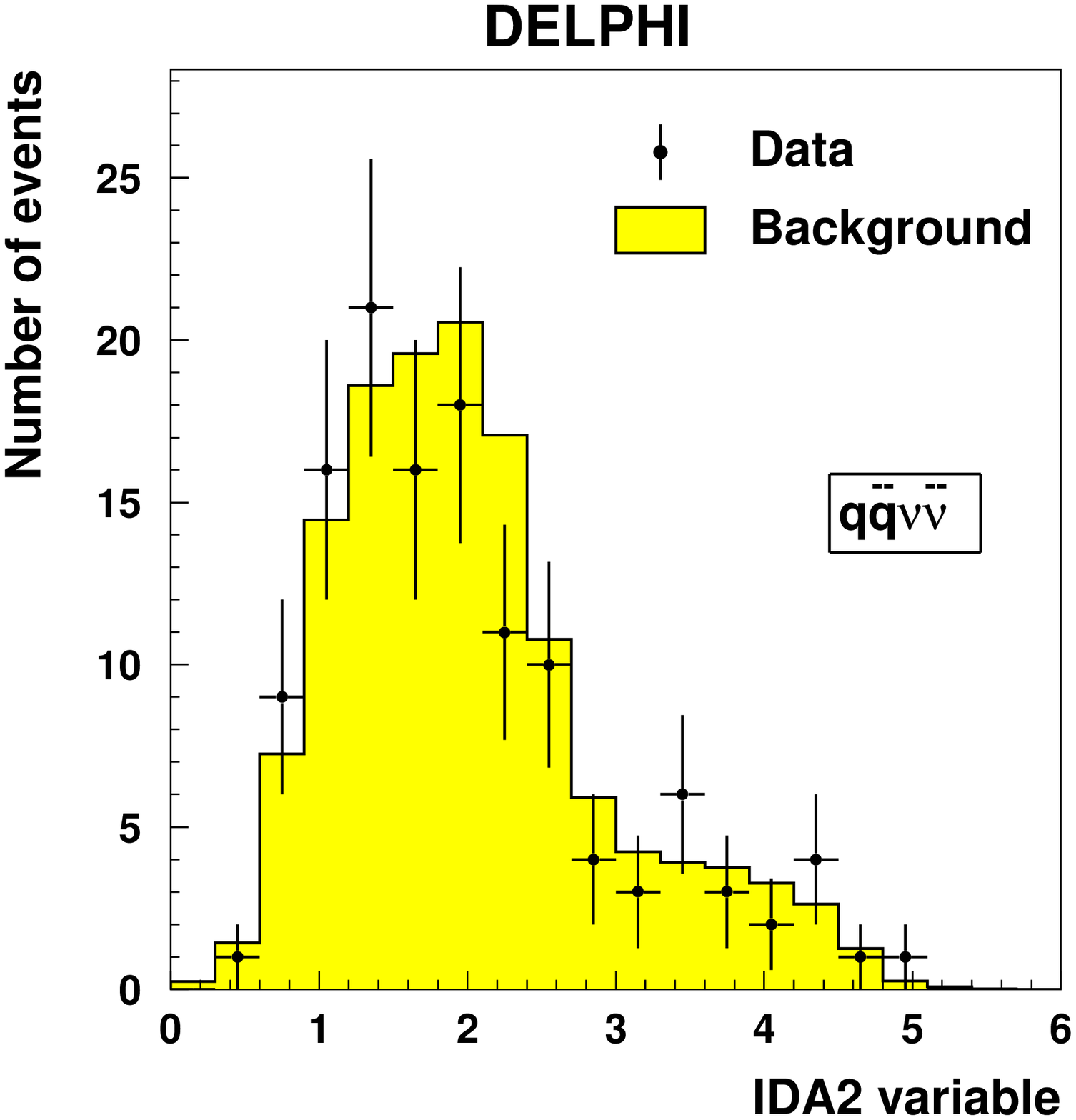}} 
\subfigure{ \includegraphics[width=.4\textwidth]{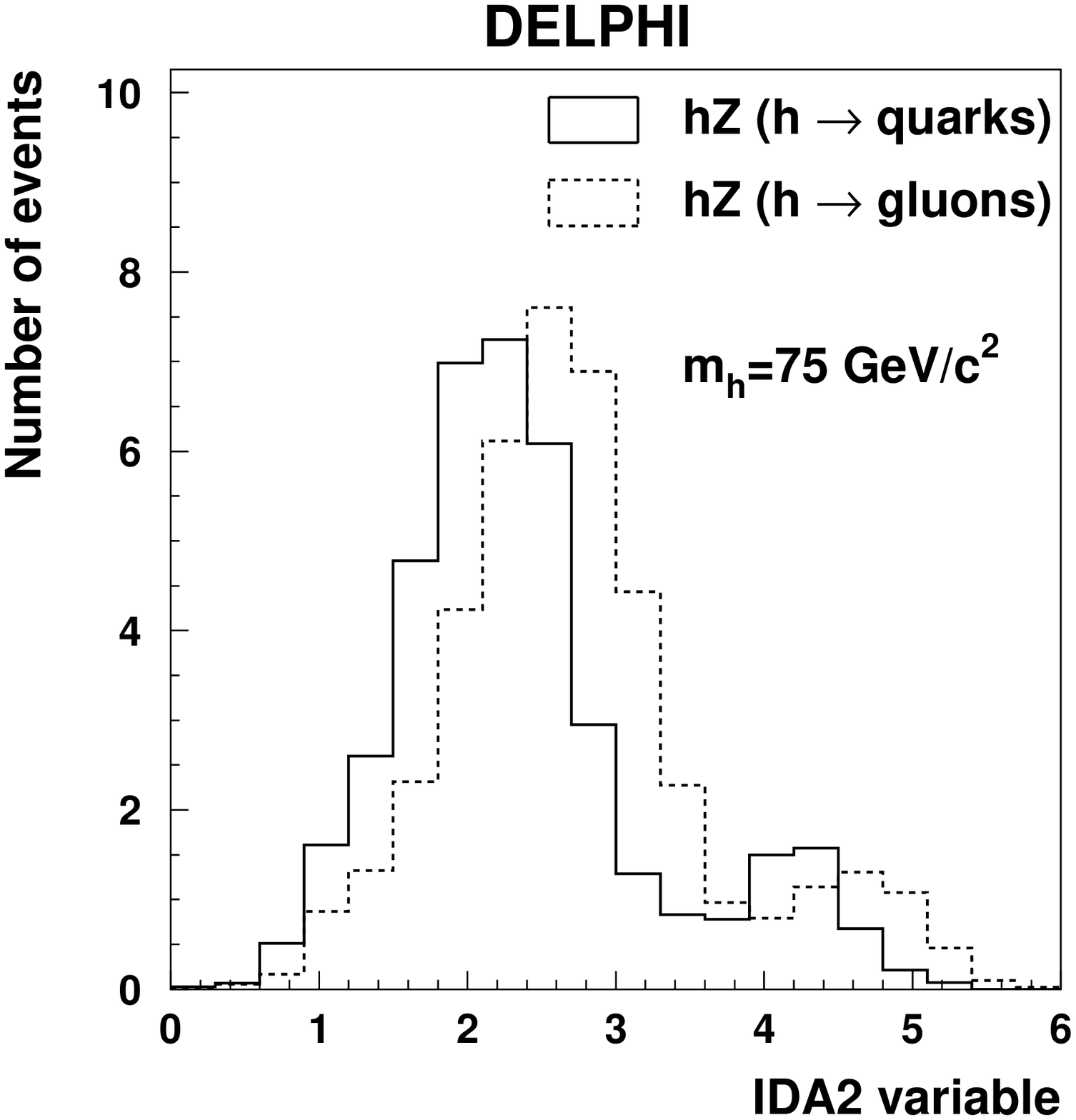}} \\
\vspace{-0.75cm}
\subfigure{ \includegraphics[width=.4\textwidth]{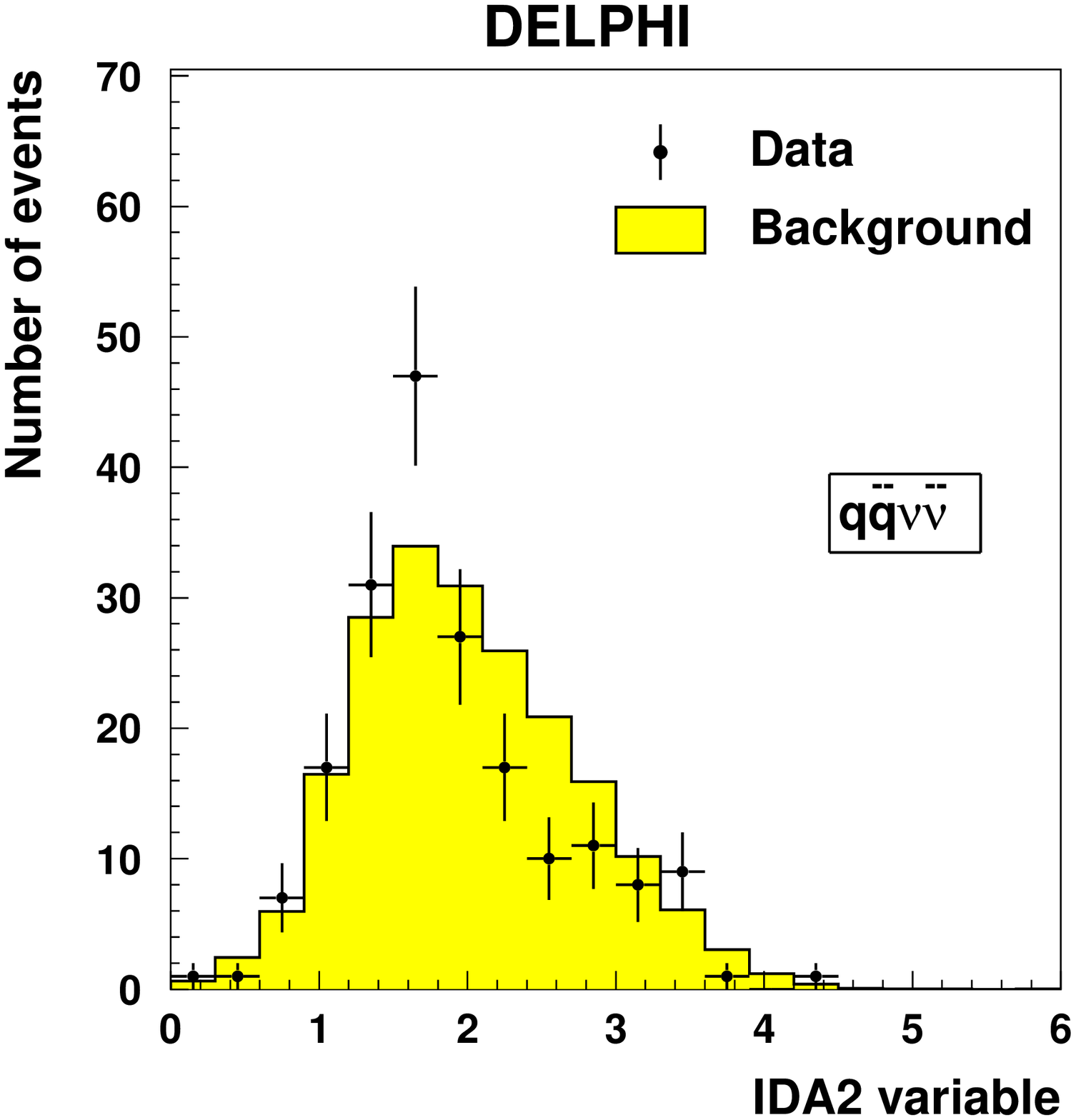}} 
\subfigure{ \includegraphics[width=.4\textwidth]{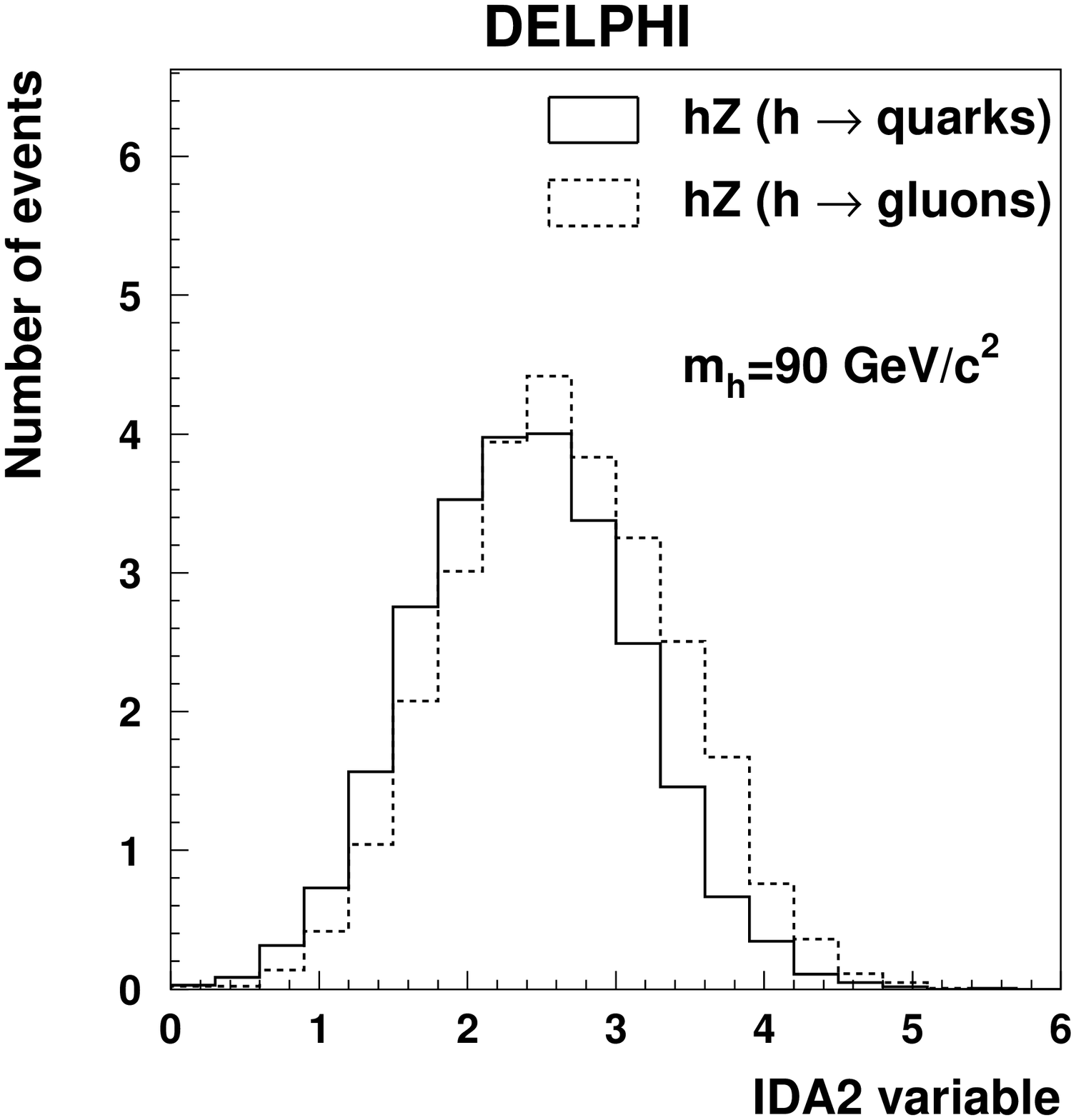}} \\
\vspace{-0.75cm}
\subfigure{ \includegraphics[width=.4\textwidth]{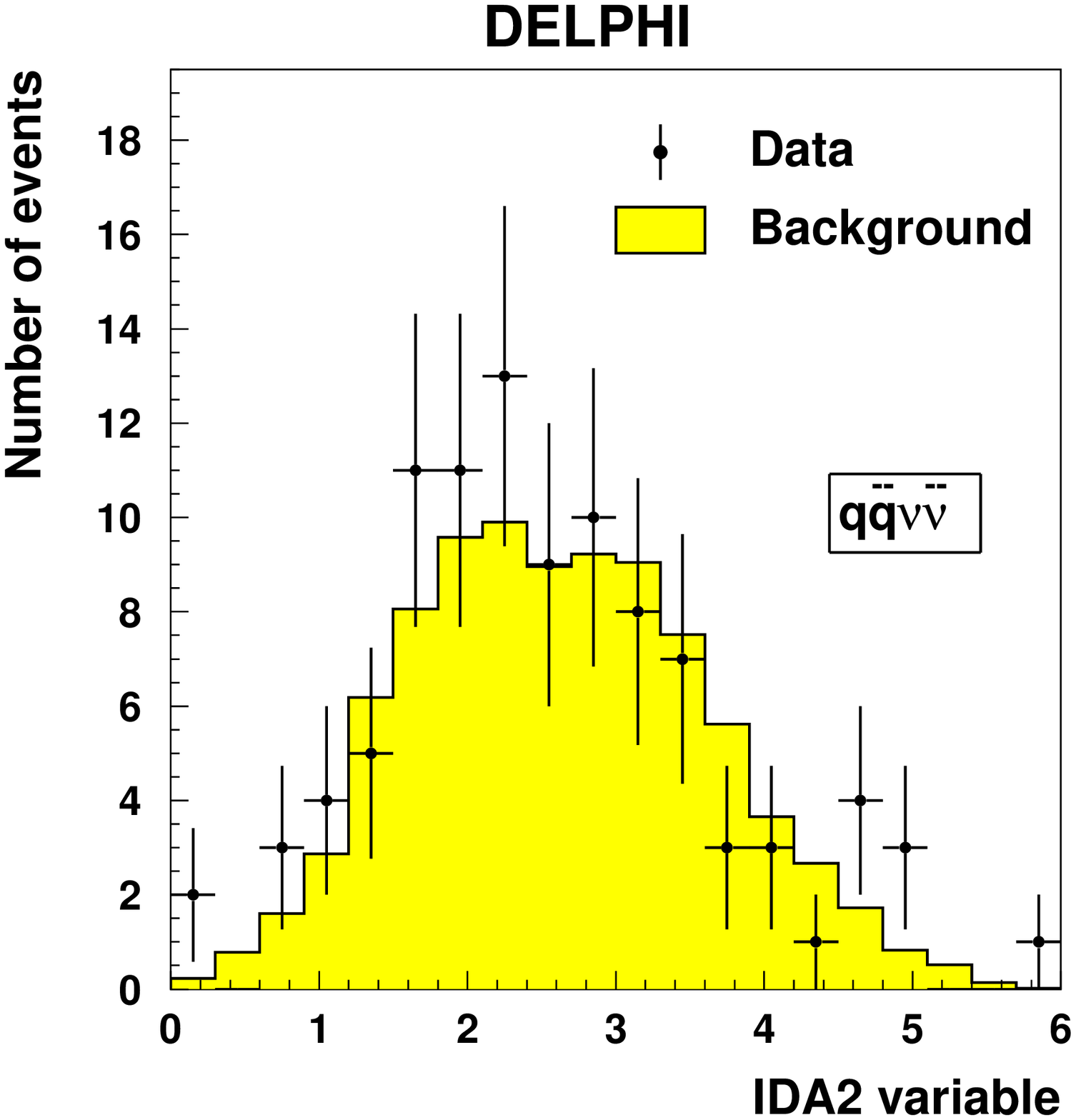}} 
\subfigure{ \includegraphics[width=.4\textwidth]{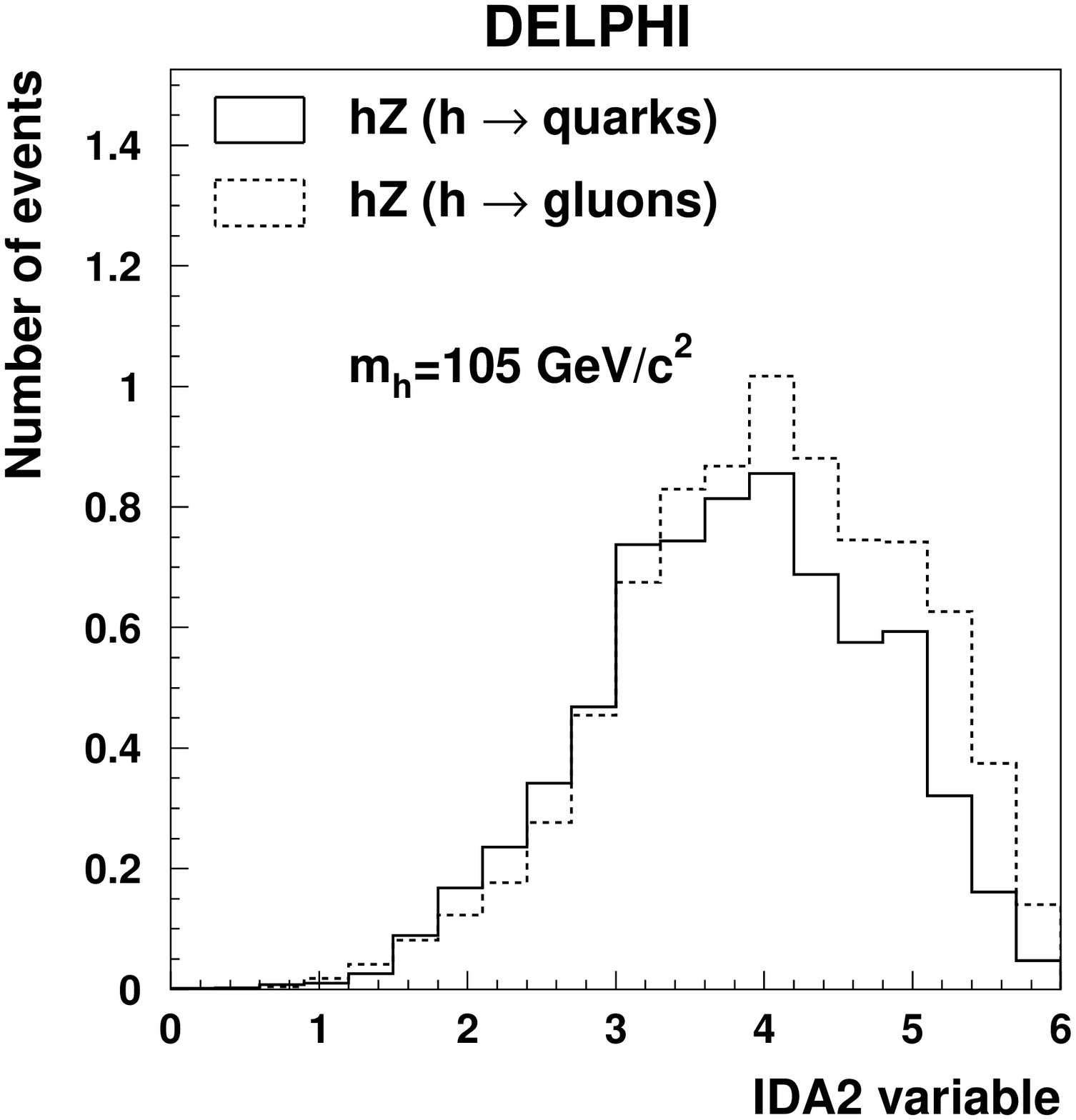}} \\
\end{center}
\vspace{-2em}
\caption{hZ search, final state with jets and missing energy, high mass search.
         Comparison between data and simulation for the distributions of IDA2, after
         all selections. The distributions correspond to Higgs boson mass hypotheses of 75, 90 and 
         105~\massunit\ from top to bottom, respectively.
         All data from 189 to 209 GeV are included. The signal normalization assumes \chz=1.}
\label{fig:hzhvvhighmass}
\end{figure}

The second step discriminant variable, called IDA2, is used in the statistical interpretation of the analyses. 
For every mass hypothesis, the signal distribution is obtained from an interpolation between the two closest simulated
signal samples.
The IDA2 distributions are shown in Figure~\ref{fig:hzhvvhighmass}, for events inside the final mass window. 
This figure also shows that a better separation between background and signal is obtained for gluonic Higgs 
boson decays.

The estimation of the systematic uncertainties also follows what was done in the
search for invisible Higgs boson decays \cite{DELPHIinvhiggs}. The dominant contribution 
originates from the description of the event shape variables, which 
affects the estimation of the \qcd~background. Calibration data taken at
centre-of-mass energies close to the Z pole are used to evaluate these systematics using 
the method described in \cite{ZZpaperDelphi,DELPHIinvhiggs}. 
Compared to these, the uncertainties on the background level due to the choice of jet 
clustering algorithm and the b-tagging procedure are small and can be neglected. 

The uncertainties on the background vary from year to year due to different detector operating 
conditions, and are between 5.7\% and 12.6\% for the first mass range, between 6.2\% and 10.2\% for the second 
mass range, and between 4.0\% and 8.2\% for the third mass range. The corresponding systematic uncertainty on the 
signal efficiency is between 1 and 3\%, the largest uncertainty affecting the first mass range.

\begin{table}[htbp]
  \begin{center}
    \begin{tabular}{r | r r | r r }
      \hline
      \hline
      \mh~(\massunit) & SM(no \hZ) & observed & {\large $\epsilon_{\mathrm{hZ}}$}(\%)  & \hZ(\h \ra \bdqq)  \\
      \hline
      ~40                &   ~~5.5    &  ~~6~~~~~      &    44.1                   &   62.4   \\
      ~50                &   ~~8.7    &  ~~8~~~~~      &    40.7                   &   51.2   \\
      ~60                &   ~16.2    &  ~18~~~~~      &    39.7                   &   43.1   \\
      ~70                &   ~82.8    &  ~72~~~~~      &    43.0                   &   39.0   \\
      ~80                &   165.2    &  155~~~~~      &    45.8                   &   32.9   \\
      ~90                &   257.9    &  238~~~~~      &    50.6                   &   25.5   \\
      100                &   121.4    &  124~~~~~      &    44.6                   &   11.2   \\
      110                &   ~66.9    &  ~76~~~~~      &    42.0                   &   ~3.3   \\
     \hline
     \end{tabular}
  \end{center}
  \caption{hZ search, final state with jets and missing energy, high mass search.
           Numerical comparison between background simulation, data, and a few example signals, just before
           the second IDA iteration.
           All data from 189 to 209 GeV are included. The signal normalization assumes \chz=1.            
           The statistical uncertainty on the signal efficiency and on the background is less than 1\%.}
  \label{tab:hZ.qqnunuhighmass.DataMC}
\end{table}

\subsection{Final states with jets and isolated leptons}
\label{sec:hzqqll}

The \qqee~and \qqmumu~channels are analysed in the same way as in 
the \epem\ra~\ZZ\ and \Zgstar~cross-section measurements.
This search selects events with two identified leptons of the same flavour (electrons or muons) and of opposite 
charge, and at least two jets reconstructed with the JADE algorithm \cite{jade} (with $y_{min}=0.01$). 
Compared to~\cite{ZZpaperDelphi,ZgpaperDelphi}, cuts on the invariant mass of the leptonic and hadronic systems 
(used to enrich the \ZZ\ and \Zgstar\ samples) are not applied here, and the particle identification criteria 
are slightly modified in order to maximize the Higgs boson signal to background ratio.

The reconstructed mass of the hadronic system is used as discriminant in the statistical interpretation of the 
analyses. For every mass hypothesis, the signal distribution is obtained from an interpolation between the two 
closest simulated signal samples, and a window is defined around the nominal Higgs boson mass to maximize
the signal to background ratio.

\begin{figure}[htbp]
\begin{center}
\subfigure{ \includegraphics[width=.4\textwidth]{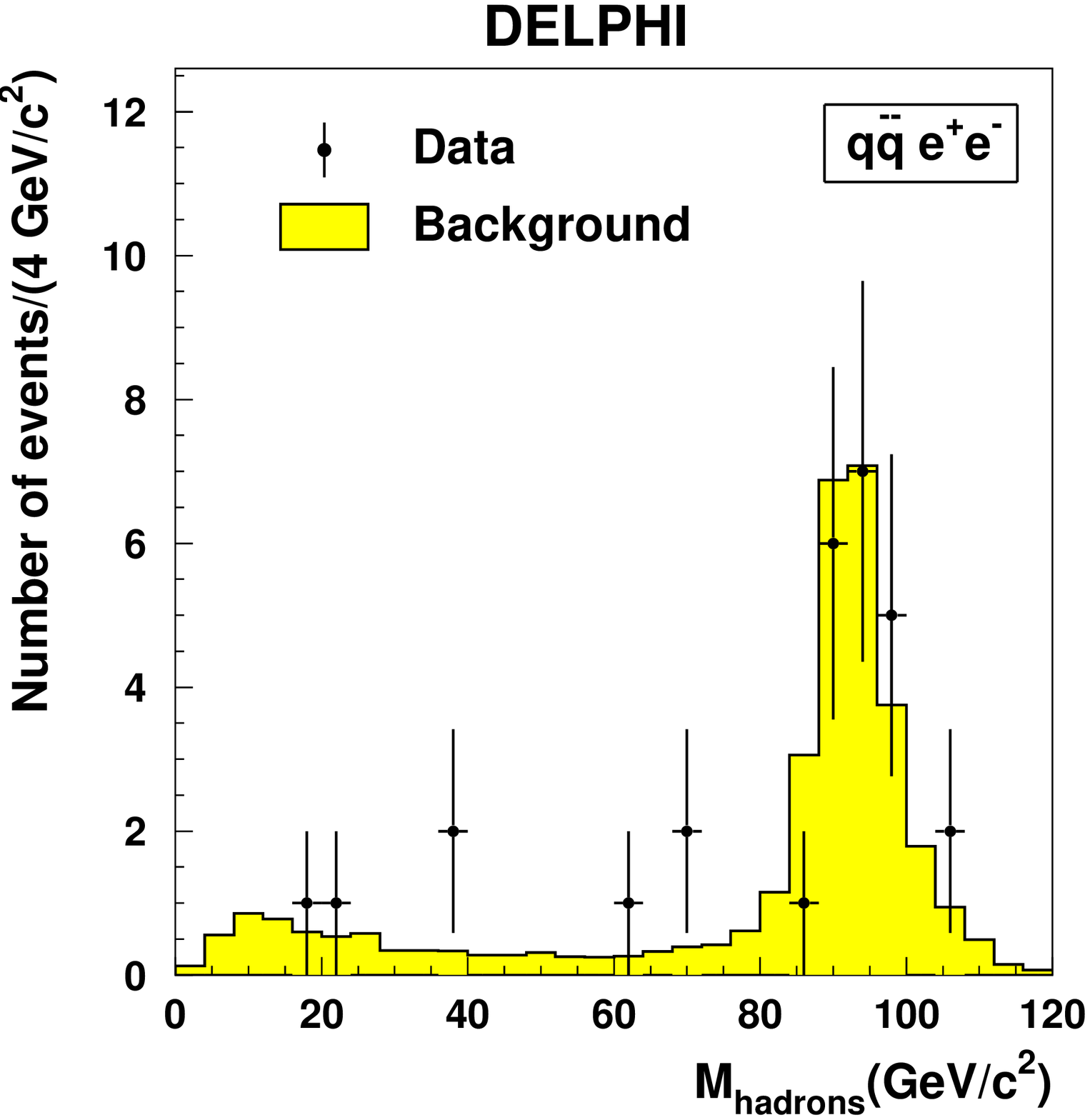}} 
\subfigure{ \includegraphics[width=.4\textwidth]{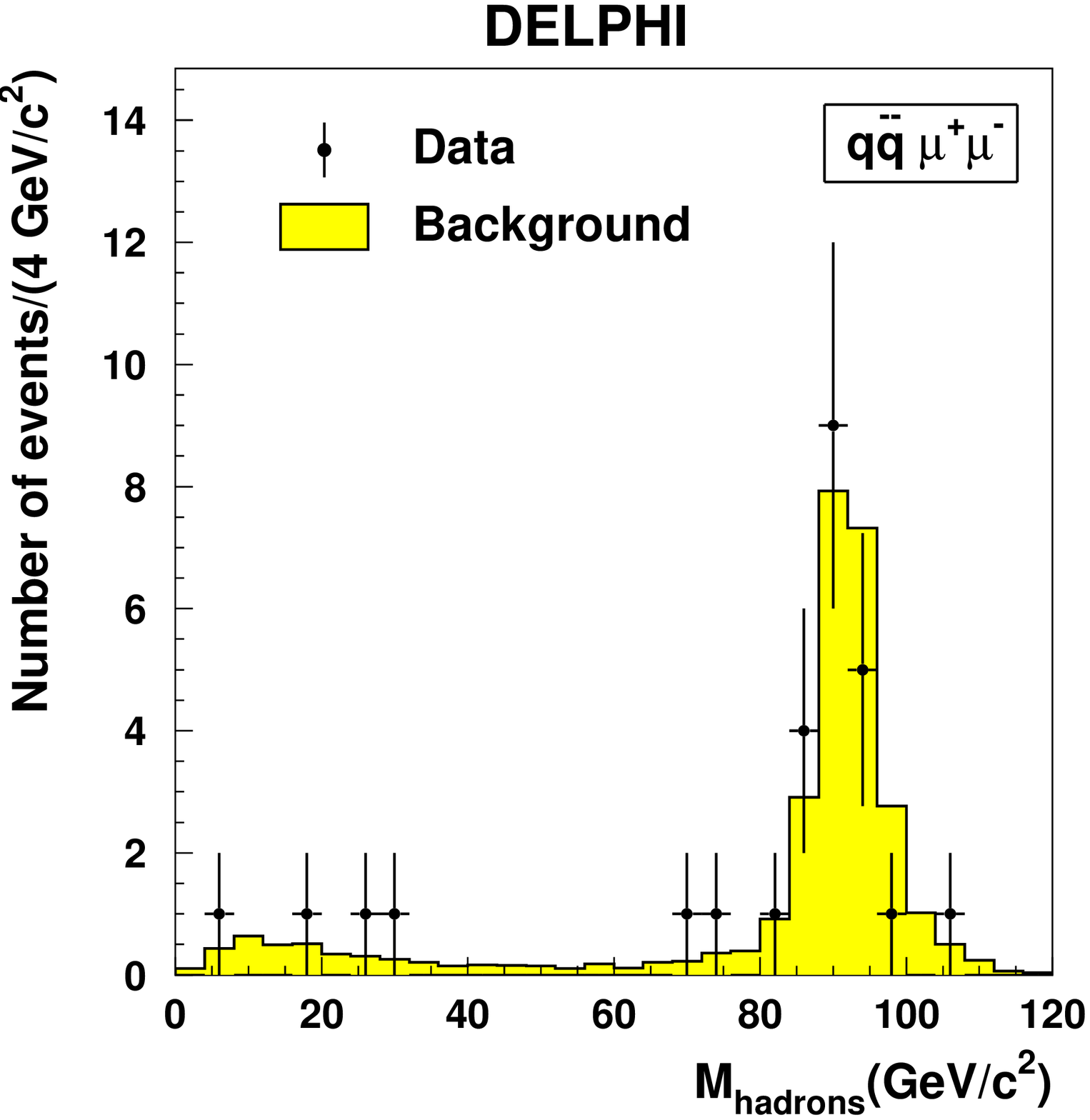}}
\end{center}
   \vspace{-2em}
   \caption{hZ search, final state with jets and isolated leptons.
           Comparison between data and simulation for the reconstructed hadronic mass distribution, when the mass of 
           the opposite lepton pair is within 15~\massunit\ of the \Z~mass. All data 
           from 189 to 209 GeV are included.}
    \label{fig:hzqqll.massplots}
\end{figure}

Distributions of the reconstructed hadronic mass are shown in Figure~\ref{fig:hzqqll.massplots}. 
Tables~\ref{tab:hZ.qqee.DataMC} and~\ref{tab:hZ.qqmumu.DataMC} display the signal efficiencies, and a
numerical comparison between data and simulation.

The main systematic uncertainties affecting the results are the uncertainty in the lepton 
identification and the signal selection efficiency. Combined with the 
statistical uncertainty from the finite size of the simulated sample, these result in a 
relative systematic uncertainty in selecting \qqee~and \qqmumu~events of 5\%. The 
uncertainty on the small residual background rate is dominated by the MC statistics and 
amounts to 15\%.

\begin{table}[htbp]
  \begin{center}
    \begin{tabular}{r | r r | r r }
      \hline
      \hline
      \mh~(\massunit)  & SM(no \hZ) & observed & {\large $\epsilon_{\mathrm{hZ}}$}(\%)  & \hZ(\h \ra \bdqq) \\
      \hline
      ~40             &  ~1.6      &    3~~~~~           &       59.8             &    14.2   \\
      ~50             &  ~1.1      &    2~~~~~           &       61.8             &    12.9   \\
      ~60             &  ~1.5      &    3~~~~~           &       63.3             &    11.4   \\
      ~70             &  ~2.3      &    4~~~~~           &       63.4             &    ~9.5   \\
      ~80             &  17.6      &   18~~~~~           &       66.1             &    ~7.9   \\
      ~90             &  23.3      &   18~~~~~           &       69.5             &    ~5.8   \\
      100             &  16.3      &   11~~~~~           &       68.2             &    ~2.8   \\
      110             &  ~1.9      &    3~~~~~           &       64.2             &    ~0.8   \\
     \hline
     \end{tabular}
  \end{center}
  \vspace{-1em}
  \caption{hZ search, final state with jets and isolated electrons.
           Numerical comparison between background simulation, data, 
           and a few example signals, after the mass cuts that maximize the separating power for each 
           Higgs boson mass hypothesis.
           All data from 189 to 209 GeV are included. The signal normalization assumes \chz=1.}     
  \label{tab:hZ.qqee.DataMC}
\end{table}

\begin{table}[htbp]
  \begin{center}
    \begin{tabular}{r | r r | r r }
      \hline
      \hline
      \mh~(\massunit)  & SM(no \hZ) & observed & {\large $\epsilon_{\mathrm{hZ}}$}(\%)  & \hZ(\h \ra \bdqq) \\
      \hline
      ~40              &   ~1.5    &   ~3~~~~~           &    77.1                  &   18.1    \\
      ~50              &   ~2.3    &   ~4~~~~~           &    77.5                  &   16.1    \\
      ~60              &   ~2.0    &   ~3~~~~~           &    78.5                  &   14.1    \\
      ~70              &   ~3.3    &   ~3~~~~~           &    76.8                  &   11.5    \\
      ~80              &   20.0    &   16~~~~~           &    78.0                  &   ~9.3    \\
      ~90              &   25.6    &   20~~~~~           &    80.6                  &   ~6.7    \\
      100              &   17.8    &   18~~~~~           &    81.1                  &   ~3.3    \\
      110              &   ~3.1    &   ~2~~~~~           &    76.8                  &   ~0.9    \\
     \hline
     \end{tabular}
  \end{center}
  \vspace{-1em}
  \caption{hZ search, final state with jets and isolated muons.
           Numerical comparison between data, background simulation,
           and a few example signals, after the mass cuts that maximize the separating power for each 
           Higgs boson mass hypothesis.
           All data from 189 to 209 GeV are included. The signal normalization assumes \chz=1.}
  \label{tab:hZ.qqmumu.DataMC}
\end{table}

\section{Results}
\label{sec:results}

The hypothesis of Higgs boson production is tested in this section, using the results from the analyses described above.
The definition of the reference cross-sections, the statistical procedure, and the scope and limitations of the results
are discussed in Section~\ref{sec:generalstrategy}. The results are represented graphically in this paper, but explicit
exclusion values can be obtained from the Collaboration on request.

\subsection{Excluded cross-sections for hA production}
\label{sec:hAresults}

For each (\mh,\mA) hypothesis, the $\mathrm{\chi^2(\mh,\mA)}$ distributions for background simulation, data and
signal (see Figure \ref{fig:hA.finaldiscriminant}) are used as input to the statistical analysis.

The three analysis streams are combined as follows. The high-thrust stream is independent from the three-jet and 
four-jet streams, and is statistically combined with each of them in turn. The resulting combinations partly overlap,
and cannot be added as independent channels. The ambiguity is solved at each mass point by choosing the combination 
that provides the strongest expected exclusion.

\begin{figure}[p]
  \begin{center}
     \includegraphics[width=.85\textwidth]{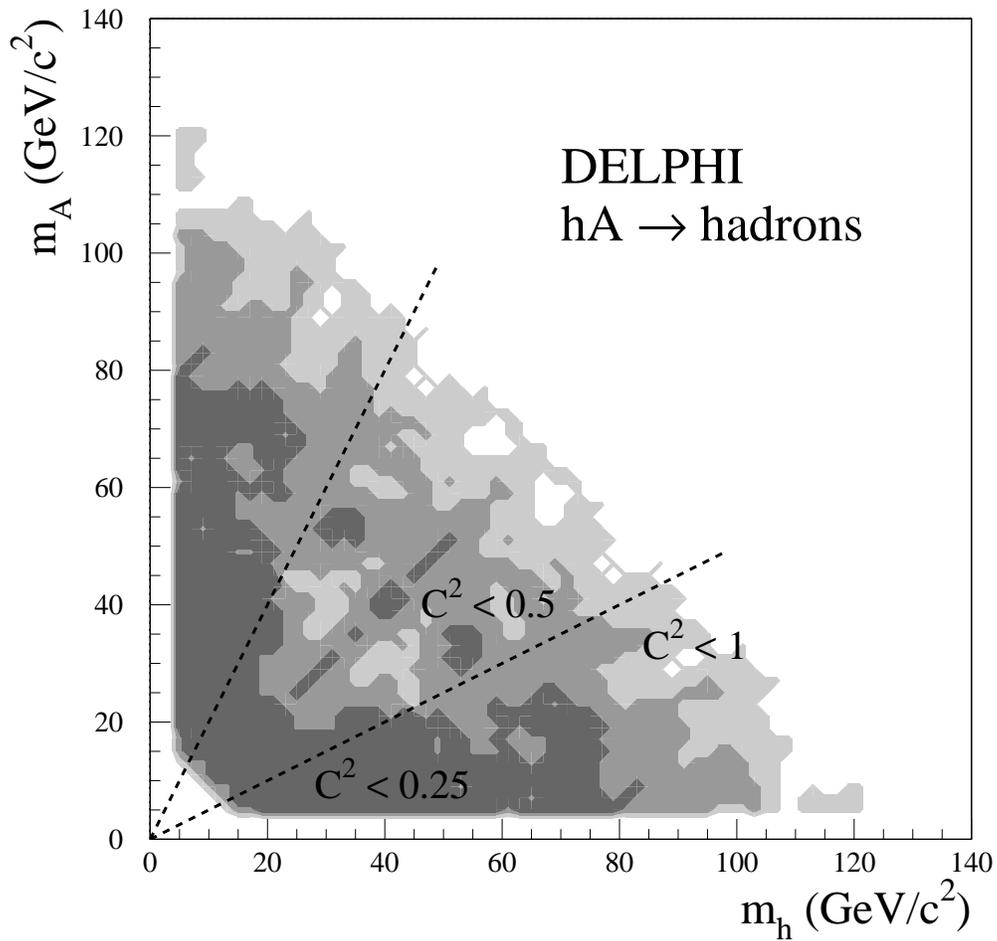}
     \vspace{-1em}
     \caption{Exclusion regions in the (\mh,\mA)-plane, for 0.5 $\leq$ \cha\ $\leq$ 1,
             0.25 $\leq$ \cha\ $\leq$ 0.5, and \cha\ $\leq$ 0.25 from light to dark grey. The dashed lines indicate the 
              borders where the h~\ra~AA decay opens kinematically. The exclusions are given at the 95\% CL.}
     \label{fig:hAresults.2dplot}
   \end{center}
\end{figure}

The mass domain is scanned from \mh,\mA$>$4 \massunit\ to the kinematic limit, in mass steps of 1 \massunit\ in 
both directions. At each mass point, the confidence level in the background hypothesis is computed, and the value 
of \cha\ that is excluded with a confidence level of exactly 95\% is determined.
No evidence for signal is observed. Figure \ref{fig:hAresults.2dplot} displays regions 
excluded in the (\mh,\mA)-plane, for a few fixed values of \cha. 
For comparable Higgs boson masses and maximal production cross-section (\cha=1), the combinations 
with \mh+\mA\ below $\sim$140 \massunit~are excluded. When 
one of the Higgs bosons is very light, the mass of the other boson is constrained to 
be either below 4 \massunit\ (the \hA~search threshold), or above 108 \massunit, still when \cha=1. 
The three-jet and high-thrust analysis streams allow a significant portion of the mass plane to be excluded 
even when \cha\ is as small as~0.25.

In most regions of the mass domain, the present search has high background and relatively low signal 
(see Figure~\ref{fig:hA.finaldiscriminant}). The exclusions are thus very sensitive to fluctuations in the data, 
which explains the irregularities in the exclusion regions.

The above results are valid in the case of direct Higgs boson decays into hadrons. 
When the intermediate h$\rightarrow$AA decay mode opens (the corresponding kinematic 
domain is delimited by the dashed lines in Figure~\ref{fig:hAresults.2dplot}), 
the six-parton final state may become dominant. The sensitivity of the present 
analyses to this topology has not been evaluated.

As stated in Section \ref{sec:generalstrategy}, the hA exclusions are valid independently of the 
parity properties of h and A, and in particular may be interpreted in models with additional Higgs doublets or singlets, or 
in the context of a CP-violating Higgs sector.

\subsection{Excluded cross-sections for hZ production}
\label{sec:hZresults}

For each \mh\ hypothesis, the discriminant variables provided by the analysis channels
described in Section~\ref{sec:hZproduction} are used as input to the statistical analysis.
For masses below  40~\massunit, these are the ${\mathrm{\chi^2}}(\mh,\mZ)$ distributions from the
hadronic channel (like in Figure~\ref{fig:hA.finaldiscriminant}), and the mass spectra from the low mass missing energy channel (see 
Figure~\ref{fig:hz.hvvlowmass}).
Above  40~\massunit, the hadronic channel provides distributions of ln(P$\mathrm{_{Higgs}}(\mh))$
(Figure~\ref{fig:hZqqqq}), the high mass missing energy search provides distributions of the second step IDA output, 
IDA2 (Figure~\ref{fig:hzhvvhighmass}), and the channel with jets and leptons provides the hadronic mass distributions.

The Higgs boson mass domain is scanned from \mh$\geq$4~\massunit\ to the kinematic limit. At each tested mass, the 
confidence level in the background hypothesis is computed, and the value of \chz\ excluded with a confidence level of
exactly 95\% is determined. The scan is done in steps of 1~\massunit\ over 
most of the range, and in steps of 100 $\mathrm{MeV/c^2}$ when the excluded value of \chz\ approaches 1.
Cross-section limits are computed for Higgs boson decays into either gluon or quark pairs, and 
the flavour-independent result at each mass is given by the weakest expected exclusion. 

\begin{figure}[htbp]
\begin{center}
  \includegraphics[width=.85\textwidth]{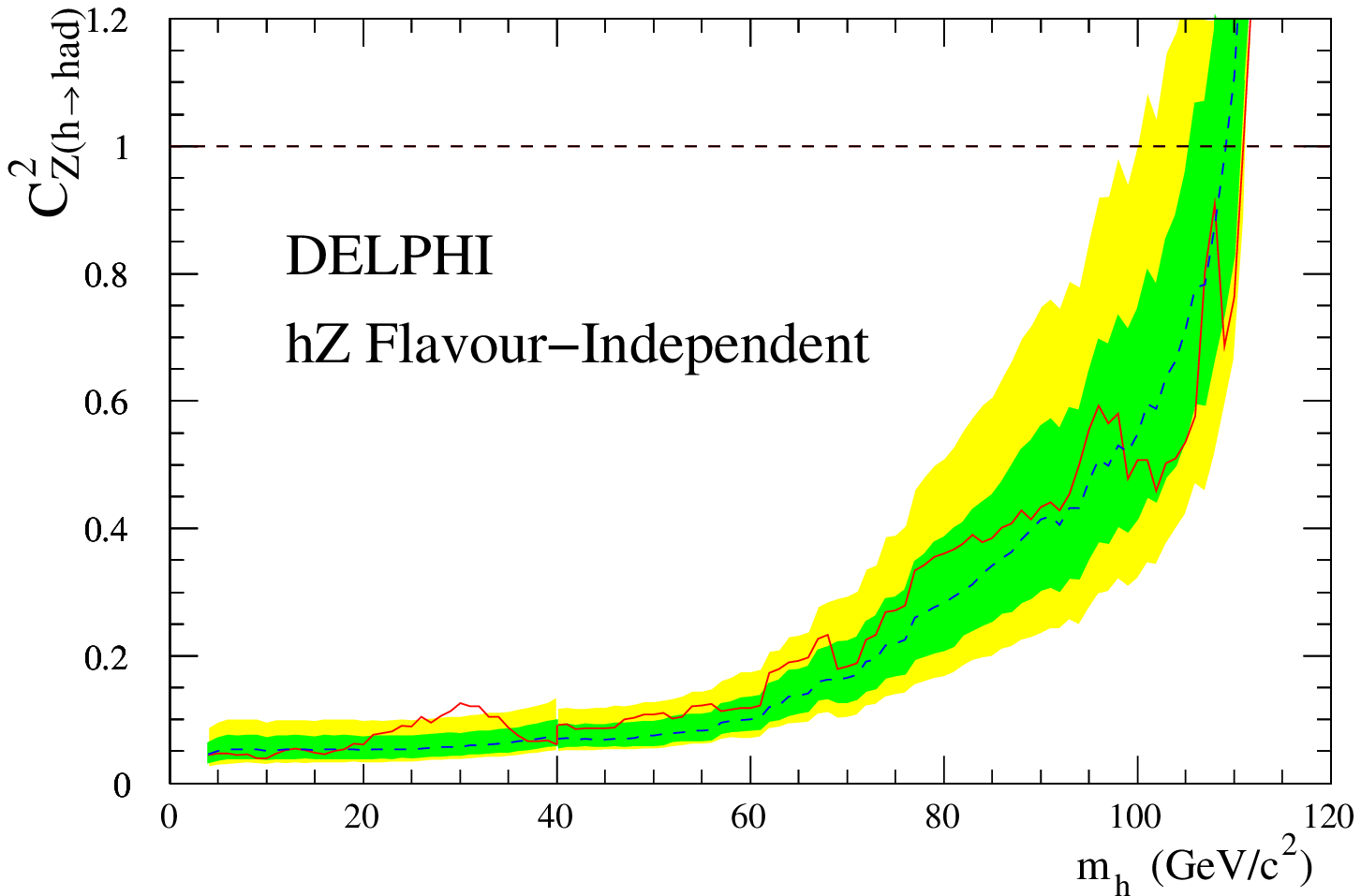}
\end{center}
\vspace{-1em}
\caption{Excluded values of \chz\ as a function of the Higgs boson mass, assuming direct hadronic Higgs boson decays.
The solid curve is the observed exclusion, the median exclusion expected in 
the absence of signal is shown by the dashed curve, and the bands represent the 68\% and 95\% confidence intervals.}
\label{fig:hzresults.csexcl}
~~\vspace{1cm}
  \begin{center}
    \includegraphics[width=.85\textwidth]{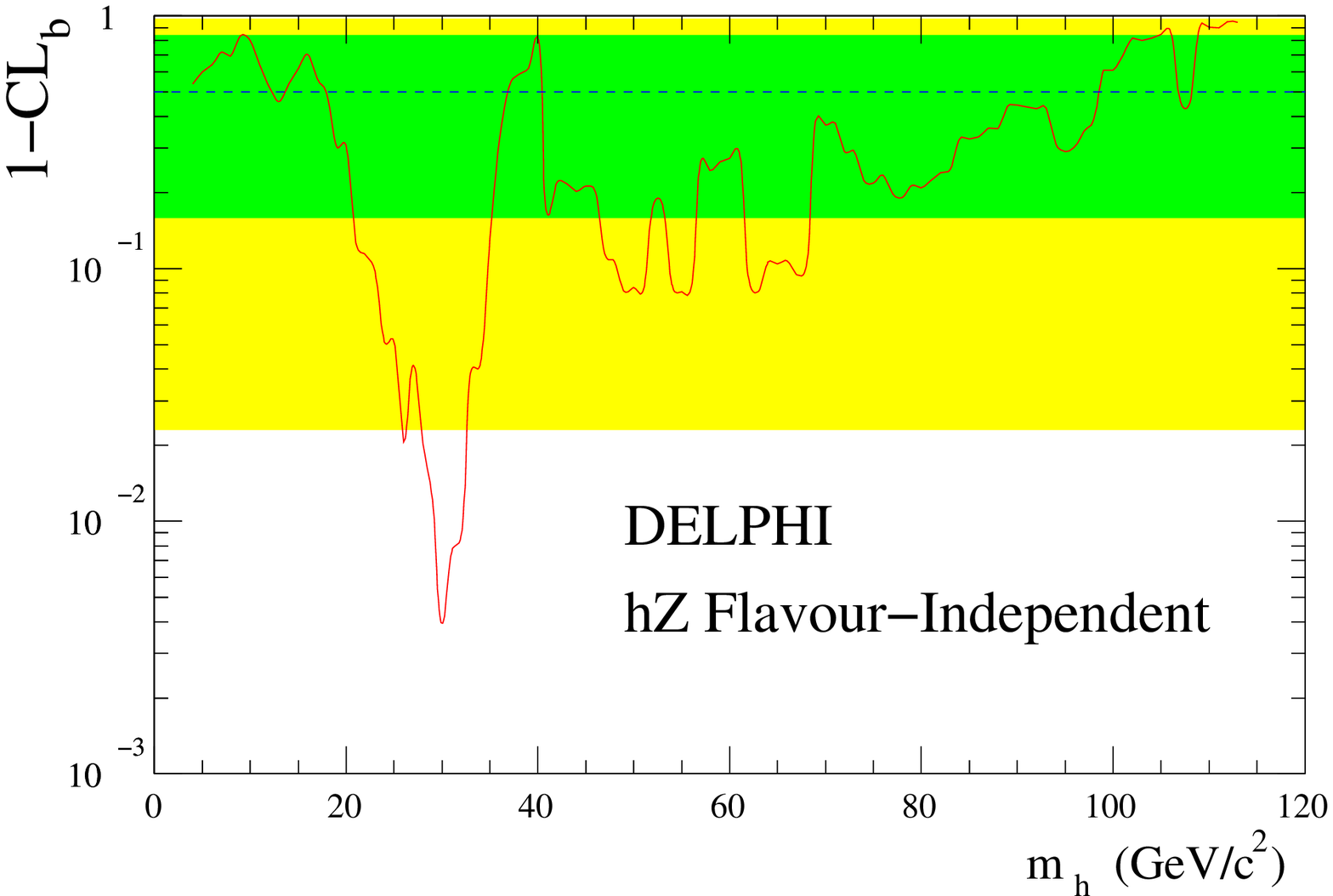}
\vspace{-1em}
\caption{Confidence level in the background hypothesis as a function of the Higgs boson mass. 
The solid curve is the observed confidence, the median expectation
from background-only experiments is shown by the dashed line, and the bands represent the 68\% and 95\% confidence intervals.}
\label{fig:hzresults.clb}
  \end{center}
\end{figure}

Figure \ref{fig:hzresults.csexcl} shows the excluded values \chz\ as a function of the Higgs bosons mass. 
Observed and expected limits agree well over a wide range of masses. The largest discrepancy occurs for a 
Higgs boson mass close to 30 \massunit~where 
an excess of 2.65 $\sigma$ is observed (the excess is around 2.5 $\sigma$ when the result is averaged 
over the mass resolution at that particular Higgs boson mass hypothesis). The excess is shared by 
two channels: the missing energy channel contributes around 2 $\sigma$, and the low-mass fully 
hadronic channel contributes around 1.5 $\sigma$. The confidence level in the background 
hypothesis is shown as a function of \mh\ in Figure \ref{fig:hzresults.clb}. The confidence level in the background 
reaches a few 10$^{-3}$ around \mh=30~\massunit, but it should be remembered that given the large number of tested 
mass hypotheses, the probability that such an excess occurs at at least one mass is much larger.

Except for signal hypotheses around \mh\ = 107 \massunit, there is a deficit of data in the hypothesis 
range 100 $\leq$ \mh\ $\leq$ 115 \massunit. The discrepancy is at the level of 1 to 1.5~$\sigma$, as can be seen in
Figure \ref{fig:hzresults.clb}. The exclusions are thus stronger than expected in this region.
Table \ref{tab:delphi.excluded.Higgsmass} gives the mass limits for hadronic Higgs boson decays
when \chz=1, and for comparison the results achieved in the SM Higgs boson searches \cite{DELPHI.SMHiggsPaper}.

\begin{table}[htbp]
 \begin{center}
\begin{tabular}{ l | c  c }
\hline     
\hline     
              ~~~                                &     expected limit   &   observed limit   \\
                                                 &    (\massunit)  &  (\massunit)  \\
   \hline                                                 
   DELPHI (\h \ra \bdqq)                         &       108.0   &      110.6      \\
   DELPHI (\h \ra \bdgg)                         &       109.2   &      111.0      \\
   DELPHI (SM decay)                             &       113.3   &      114.1      \\
  \hline      
\end{tabular}
\caption{The expected and observed 95\%~CL lower limits (in \massunit) on the mass of 
         a hadronically decaying Higgs boson, assuming its cross-section is identical 
         to that in the SM. The results of the SM  Higgs boson search are also given 
         for comparison.}
\label{tab:delphi.excluded.Higgsmass}
\end{center}
\end{table}

\section{Summary}
\label{sec:summary}

Searches for hadronic Higgs boson decays have been performed, exploiting 
the LEP2 collision data recorded by DELPHI. No evidence of the presence of a signal was found in either the
\epem\ra\hZ\ or the \epem\ra\hA\ process, in a mass range extending from 4 \massunit\ to the kinematic 
limit.

For the \hZ~process, a cross-section larger than 15\% of the SM value (i.e \chz$>$0.15) is excluded from
\mh=4 to 60~\massunit. Under the assumption of 
a production cross-section equal to that of the Standard Model, \mh$<$110.6~\massunit\ 
is excluded at the 95\%~CL, while a limit of 108~\massunit\ is expected on average from background-only 
experiments. The cross-section limit is weaker than expected around \mh=30~\massunit.

A large part of the mass domain available for the hA process was explored as well. 
In the case of full production strength and 100\% branching fraction into hadrons (i.e. \cha=1),
the excluded region extends roughly up to \mh,\mA=108~\massunit~for the heavier 
Higgs boson, when the lighter one has a mass below 10~\massunit. When both masses 
are equal,  \mh,\mA$<$70 \massunit~is excluded. Even for reduced cross-sections, 
significant portions of the mass plane remain excluded.

\subsection*{Acknowledgements}
\vskip 3 mm
 We are greatly indebted to our technical 
collaborators, to the members of the CERN-SL Division for the excellent 
performance of the LEP collider, and to the funding agencies for their
support in building and operating the DELPHI detector.\\
We acknowledge in particular the support of \\
Austrian Federal Ministry of Education, Science and Culture,
GZ 616.364/2-III/2a/98, \\
FNRS--FWO, Flanders Institute to encourage scientific and technological 
research in the industry (IWT), Belgium,  \\
FINEP, CNPq, CAPES, FUJB and FAPERJ, Brazil, \\
Czech Ministry of Industry and Trade, GA CR 202/99/1362,\\
Commission of the European Communities (DG XII), \\
Direction des Sciences de la Mati$\grave{\mbox{\rm e}}$re, CEA, France, \\
Bundesministerium f$\ddot{\mbox{\rm u}}$r Bildung, Wissenschaft, Forschung 
und Technologie, Germany,\\
General Secretariat for Research and Technology, Greece, \\
National Science Foundation (NWO) and Foundation for Research on Matter (FOM),
The Netherlands, \\
Norwegian Research Council,  \\
State Committee for Scientific Research, Poland, SPUB-M/CERN/PO3/DZ296/2000,
SPUB-M/CERN/PO3/DZ297/2000, 2P03B 104 19 and 2P03B 69 23(2002-2004)\\
FCT - Funda\c{c}\~ao para a Ci\^encia e Tecnologia, Portugal, \\
Vedecka grantova agentura MS SR, Slovakia, Nr. 95/5195/134, \\
Ministry of Science and Technology of the Republic of Slovenia, \\
CICYT, Spain, AEN99-0950 and AEN99-0761,  \\
The Swedish Research Council,      \\
Particle Physics and Astronomy Research Council, UK, \\
Department of Energy, USA, DE-FG02-01ER41155. \\
EEC RTN contract HPRN-CT-00292-2002. \\




\end{document}